\documentclass[twocolumn]{aastex701}

\usepackage{amsmath}
\usepackage{amssymb}
\usepackage{graphicx}
\usepackage{color}
\usepackage{comment}

\usepackage{savesym}
\savesymbol{tablenum}
\usepackage{siunitx}
\usepackage{xspace}
\restoresymbol{SIX}{tablenum}

\definecolor{myblue}{RGB}{40,40,250}

\newcommand\sgquote[1]{\textquoteleft #1\textquoteright}

\newcommand\partialdiff[1]{\frac{\partial}{\partial #1}}

\newcommand\partiallogdiff[1]{\textrm{d}\mathrm{ln} #1/\textrm{d} \mathrm{ln}\, r}

\newcommand\Sigmagas{\Sigma_\textrm{g}}

\graphicspath{{./}{./figures/}{./Plots_Setup/}}

\shorttitle{}
\shortauthors{Kurtovic, G\'arate, Pinilla, et al.}

\begin{document}

\title{The ALMA Survey of Gas Evolution of PROtoplanetary Disks (AGE-PRO): \\ VI. Comparison of Dust Evolution Models to AGE-PRO Observations}

%\author{Kurtovic, G\'arate, Pinilla, et al., }

%\correspondingauthor{Nicolas Kurtovic}

\author[0000-0002-2358-4796]{Nicol\'as T. Kurtovic}
\affiliation{Max Planck Institute for Extraterrestrial Physics, 
Giessenbachstrasse 1, D-85748 Garching, Germany}
\affiliation{Max-Planck-Institut fur Astronomie (MPIA), 
Konigstuhl 17, 69117 Heidelberg, Germany}
\email{kurtovic@mpe.mpg.de}

\author[0000-0001-6802-834X]{Mat\'ias G\'arate}
\affiliation{Max-Planck-Institut fur Astronomie (MPIA), 
Konigstuhl 17, 69117 Heidelberg, Germany}
\email{}

\author[0000-0001-8764-1780]{Paola Pinilla}
\affiliation{Mullard Space Science Laboratory, University College London, 
Holmbury St Mary, Dorking, Surrey RH5 6NT, UK}
\affiliation{Max-Planck-Institut fur Astronomie (MPIA), 
Konigstuhl 17, 69117 Heidelberg, Germany}
\email{}

\author[0000-0002-0661-7517]{Ke Zhang}
\affiliation{Department of Astronomy, University of Wisconsin-Madison, 
475 N Charter St, Madison, WI 53706, USA}
\email{}

\author[0000-0003-4853-5736]{Giovanni P. Rosotti}
\affiliation{Dipartimento di Fisica, Università degli Studi di Milano, Via Celoria 16, I-20133 Milano, Italy}
\email{}

\author[0009-0004-8091-5055]{Rossella Anania}
\affiliation{Dipartimento di Fisica, Università degli Studi di Milano, Via Celoria 16, I-20133 Milano, Italy}
\email{}

\author[0000-0001-7962-1683]{Ilaria Pascucci}
\affiliation{Lunar and Planetary Laboratory, the University of Arizona, Tucson, AZ 85721, USA}
\email{}

\author[0000-0002-1103-3225]{Beno\^it Tabone }
\affiliation{Université Paris-Saclay, CNRS, Institut d'Astrophysique Spatiale, 91405 Orsay, France}
\email{}

\author[0000-0002-8623-9703]{Leon Trapman}
\affiliation{Department of Astronomy, University of Wisconsin-Madison, 
475 N Charter St, Madison, WI 53706, USA}
\email{}

\author[0000-0003-0777-7392]{Dingshan Deng}
\affiliation{Lunar and Planetary Laboratory, the University of Arizona, Tucson, AZ 85721, USA}
\email{}

\author[0000-0002-4147-3846]{Miguel Vioque}
\affiliation{European Southern Observatory, Karl-Schwarzschild-Str. 2, 85748 Garching bei München, Germany}
\affiliation{Joint ALMA Observatory, Avenida Alonso de C\'ordova 3107, Vitacura, Santiago, Chile}
\email{}

\author[0000-0003-2251-0602]{John Carpenter}
\affiliation{Joint ALMA Observatory, Avenida Alonso de C\'ordova 3107, Vitacura, Santiago, Chile}
\email{}

\author[0000-0002-2828-1153]{Lucas A. Cieza}
\affiliation{Instituto de Estudios Astrofísicos, Universidad Diego Portales, Av. Ejercito 441, Santiago, Chile}
\affiliation{Millennium Nucleus on Young Exoplanets and their Moons (YEMS), Chile}
\email{}

\author[0000-0002-1199-9564]{Laura M. P\'erez}
\affiliation{Departamento de Astronom\'ia, Universidad de Chile, Camino El Observatorio 1515, Las Condes, Santiago, Chile}
\email{}

\author[0000-0002-7238-2306]{Carolina Agurto-Gangas}
\affiliation{Departamento de Astronom\'ia, Universidad de Chile, Camino El Observatorio 1515, Las Condes, Santiago, Chile}
\email{}

\author[0000-0002-5991-8073]{Anibal Sierra}
\affiliation{Mullard Space Science Laboratory, University College London, 
Holmbury St Mary, Dorking, Surrey RH5 6NT, UK}
\affiliation{Departamento de Astronom\'ia, Universidad de Chile, Camino El Observatorio 1515, Las Condes, Santiago, Chile}
\email{}

\author[0000-0003-3573-8163]{Dary A. Ru\'iz-Rodr\'iguez}
\affiliation{National Radio Astronomy Observatory, 520 Edgemont Rd., Charlottesville, VA 22903, USA}
\email{}

\author[0000-0002-1575-680X]{James Miley}
\affiliation{Departamento de Física, Universidad de Santiago de Chile, Av. Victor Jara 3659, Santiago, Chile}
\affiliation{Millennium Nucleus on Young Exoplanets and their Moons (YEMS), Chile}
\affiliation{Center for Interdisciplinary Research in Astrophysics and Space Exploration (CIRAS), Universidad de Santiago de Chile, Chile}
\email{}

\author[0000-0003-4907-189X]{Camilo Gonz\'alez-Ruilova}
\affiliation{Instituto de Estudios Astrof\'isicos, Universidad Diego Portales, Av. Ejercito 441, Santiago, Chile}
\affiliation{Millennium Nucleus on Young Exoplanets and their Moons (YEMS), Chile}
\affiliation{Center for Interdisciplinary Research in Astrophysics and Space Exploration (CIRAS), Universidad de Santiago de Chile, Chile}
\email{}

\author[0000-0001-9961-8203]{Estephani Torres-Villanueva}
\affiliation{Department of Astronomy, University of Wisconsin-Madison, 
475 N Charter St, Madison, WI 53706, USA}
\email{}

\author[0000-0002-6946-6787]{Aleksandra Kuznetsova}
\affiliation{Center for Computational Astrophysics, Flatiron Institute, 162 Fifth Ave., New York, New York, 10025}
\affiliation{Department of Physics, University of Connecticut, 196A Auditorium Road, Unit 3046, Storrs, CT 06269, USA}
\email{}

\begin{abstract}

The potential for planet formation of a circumstellar disk depends on the dust and gas reservoirs, which evolve as a function of the disk age. The ALMA Large Program AGE-PRO has measured several disk properties across three star-forming regions of different ages, and in this study we compare the observational results to dust evolution simulations. 
Using \texttt{DustPy} for the dust evolution, and \texttt{RADMC-3D} for the radiative transfer, we ran a large grid of models spanning stellar masses of 0.25, 0.50, 0.75, and 1.0\,$M_\odot$, with different initial conditions, including: disk sizes, disk gas masses, and dust-to-gas ratio, and viscosity. 
Our models are performed assuming smooth, weakly, or strongly substructured disks, aiming to investigate if any observational trend can favor or exclude the presence of dust traps. 
The observed gas masses in the disks of the AGE-PRO sample are not reproducible with our models, which only consider viscous evolution with constant $\alpha$, suggesting that additional physical mechanisms play a role in the evolution of the gas mass of disks. 
When comparing the dust continuum emission fluxes and sizes at 1.3\,mm, we find that most of the disks in the AGE-PRO sample are consistent with simulations that have either weak or strong dust traps. The evolution of spectral index in the AGE-PRO sample is also suggestive of an unresolved population of dust traps. Future observations at high angular resolution are still needed to test several hypotheses that result from comparing the observations to our simulations, including that more massive disks in gas mass have the potential to form dust traps at larger disk radii.

\end{abstract}

\keywords{accretion, accretion disks --- hydrodynamics ---  protoplanetary disks
}

%%%%%%%%%%%%%%%%%%%%%%%%%%%%%%%%%%%%%%%%%%%%%%
\section{Introduction} \label{section_Intro}
%%%%%%%%%%%%%%%%%%%%%%%%%%%%%%%%%%%%%%%%%%%%%%

Planets are formed from the gas and dust in young circumstellar disks. The diverse properties observed in the exoplanet population, and the architecture of their planetary systems, will depend on the properties and evolution of their natal disks. Gas evolution can lead to (magneto-)hydrodynamical instabilities that could create pressure bumps in the disks, halting the dust radial drift and allowing dust particles to efficiently grow from micron-sized particles to kilometer-sized bodies \citep[see][for recent reviews]{Bae2023, Drazkowska2023, lesur2023}. An important consideration is that the radial distribution of dust and gas in disks evolve as a function of time. The evolution of the dust particles strongly depends on the distribution and evolution of the gas \citep{birnstiel2023, marelpinilla2023}. The youngest disks ($<1\,$Myr) have a larger mass reservoir, and also a higher frequency rate of mass replenishment through material infall \citep{kuffmeier2023}. 
In middle-aged disks ($\sim1-3\,$Myr), most of their disk material is concentrated in the midplane where dust particles are typically observed in ring-like substructures \citep[e.g.,][]{long2018, Andrews2020_review, villenave2020}. Towards the end of a disk lifetime ($3\sim10\,$Myr), the gas disk mass reservoir is expected to decrease to undetectable levels, as well as the dust mass, although the amount of dust mass over time strongly depends on the presence or absence of substructures in disks \citep[e.g.,][]{Pinilla2020}. Thus, by contrasting gas and dust observations of planet-forming disks at different stages of evolution to gas and dust evolution models, we can identify evolutionary patterns and better understand the potential for planet formation of a disk.

Large observational samples of planet-forming disks have revealed correlations between the properties of disks and their stellar hosts. For example, stars of larger masses typically have disks of larger continuum luminosity \citep[$M_\star-L_{\text{mm}}$ relation,][]{Ansdell2016, barenfeld2016, pascucci2016, testi2022}. Similarly, more-massive stars typically have higher mass accretion rates, which also correlates with the disk (dust) mass \citep{Manara2016_a, manara2023}. Correlations between the properties of disks have also been found, such as the size-luminosity relation, where the disk size in millimeter continuum emission positively correlates with the millimeter luminosity \citep[$R_{\text{mm}}$-$L_{\text{mm}}$,][]{Tripathi2017, hendler2020}. When studying these relations as a function of disk age, we can begin testing the mechanisms driving disk and planet formation evolution, such as viscous evolution models, MHD-disk driven evolution models, and internal and external photoevaporation \citep[e.g.,][]{Garate2021, somigliana2022, somigliana2023, Tabone2022, garate2023, toci2023}. 

An important caveat to these observational trends is the calculation of disk masses and sizes from observations at millimeter wavelengths. The gas properties are often inferred from the dust properties. When measuring the dust mass content of a disk, the results rely on assuming optically thin emission, dust opacity, and dust temperature, all of them carrying significant uncertainties. Then, obtaining the gas mass requires an additional assumption on the gas-to-dust ratio in the disk. Therefore, observationally constraining the gas mass and its distribution has remained challenging, which also limits the comparison and interpretation of planet-formation models. 
This caveat also limits the comparison with dust evolution models. As the evolution of the dust strongly depends on its coupling with the gas, key processes of dust evolution (such as drift, settling, and diffusion) depend on the gas mass of the disk, which is still unknown for the large majority of systems. 

The observations of the Atacama Large (sub-)Millimeter Array (ALMA) Large Program AGE-PRO has allowed for the measurement of the gas mass and size in a sample of 30 disks three different star forming regions (SFRs), which are representative of young, middle-aged, and evolved disks \citep{AGEPRO_I_overview}. 

These measurements provide the first opportunity to compare global disk evolution models to observations of disk gas masses and sizes, for a large sample of disks. These comparisons are relevant not only to understanding current observational trends but also to exploring which ones we might be missing, and contribute to guiding future efforts toward better characterizing planet-forming disks. 

In the present work, we investigate the role of dust traps and viscous evolution in influencing the evolution of the dust reservoir, and we compare them with observational trends across the AGE-PRO sample. In Sect.~\ref{section_Modelling}, we describe the models and the explored parameter space for the dust evolution and radiative transfer models. Sect.~\ref{section_Results} summarizes the results of our models and the comparison with the AGE-PRO observations. In Sections \ref{sect:discussion} and \ref{section_Summary} we present the discussion and conclusions of this work, respectively.

%%%%%%%%%%%%%%%%%%%%%%%%%%%%%%%%%%%%%%%%%%%%%%
\section{Dust model and Synthetic observations} \label{section_Modelling}
%%%%%%%%%%%%%%%%%%%%%%%%%%%%%%%%%%%%%%%%%%%%%%

We simulate the temporal and radial evolution of the dust size distribution in planet-forming disks for a wide parameter space, along with the relevant observable quantities, such as the flux in the millimeter continuum, the corresponding gas and dust disk size, the dust and gas masses, the dust-to-gas ratio, pebble flux into the inner disk, and the accretion rates. 
For the gas and dust evolution, we used the 1D dust evolution code \texttt{Dustpy} \footnote{\texttt{DustPy} is available on \href{https://github.com/stammler/DustPy}{github.com/stammler/DustPy}. Version 0.5.6 was used for this work.} \citep{Stammler2022_Dustpy} which is based on the disk model of \cite{Birnstiel2010}, and includes the effects of coagulation and fragmentation of multiple dust species (by dust species, we refer to dust grain of different sizes), their radial advection and diffusion, as well as the viscous evolution of the gas. 

Then, we used the radiative transfer code \texttt{RADMC-3D} \footnote{\href{https://www.ita.uni-heidelberg.de/~dullemond/software/radmc-3d/}{www.ita.uni-heidelberg.de/$\sim$dullemond/software/radmc-3d/}.} \citep{RADMC3D2012} to post-process the simulations at selected snapshots, obtaining the corresponding emission in the millimeter continuum.
We compare the properties of these radiative transfer images directly to those measured by AGE-PRO observations, as the AGE-PRO disk sizes were obtained using visibility modeling \citep{AGEPRO_X_dust_disks}, which should alleviate the effect of beam convolution. 
In the Appendix, we also describe the comparison between the true properties of the radiative transfer images, and those that are measured from images convolved with a Gaussian beam and with thermal noise, to simulate the conditions of the CLEAN images. 
In this section, we provide an overview of our workflow, and describe the relevant parameters of each code. For a detailed description of \texttt{Dustpy} and \texttt{RADMC-3D} we refer to \citet{Stammler2022_Dustpy} and \citet{RADMC3D2012}.

\subsection{Dust Evolution}

The advection, diffusion, and growth or fragmentation of dust particles are strongly determined by their Stokes number, $\mathrm{St}$, which can be understood as the dimensionless coupling time to the gas motion or, more intuitively, as the dynamical grain size. 
The Stokes number is related to the physical grain size $a$ through the following equation:
\begin{equation} \label{eq_StokesMidplane}
    \textrm{St} = \frac{\pi}{2}\frac{a\, \rho_s}{\Sigma_g} \cdot
    \begin{cases}
				1 & \lambda_\textrm{mfp}/a \geq 4/9\\
                \frac{4}{9} \frac{a}{\lambda_\textrm{mfp}} & \lambda_\textrm{mfp}/a < 4/9,
	\end{cases} 
\end{equation}
where $\rho_s$ is the grain material density, $\Sigmagas$ is the gas surface density, and $\lambda_\textrm{mfp}$ is the gas mean free path, where the latter is used to distinguish between the Epstein and Stokes I drag regimes \citep{Birnstiel2010}.

Then, the radial velocity of dust particles with a given size particle is defined through:
\begin{equation} \label{eq_dust_radial_velocity}
    v_\textrm{d} = \frac{1}{1 + \mathrm{St}^2} v_\nu -  \frac{2 \mathrm{St}}{1 + \mathrm{St}^2} \eta v_k,
\end{equation}
where the first term represents the coupling to the gas radial viscous velocity \citep{Pringle1981}, and:
\begin{equation} \label{eq_velocity_visc}
v_\nu = -\frac{3}{\Sigma_\textrm{g} \sqrt{r}}\partialdiff{r}(\nu \, \Sigma_\textrm{g} \, \sqrt{r}),
\end{equation}
with $\nu = \alpha c_s h_\textrm{g}$ being the kinematic viscosity \citep{Shakura1973}, $c_s$ is the sound speed, $h_g$ is the pressure scale height, $r$ is the radial coordinate, and $\alpha$ is the dimensionless turbulence parameter. The second term corresponds to the radial drift induced by the exchange of angular momentum between gas and dust \citep{Nakagawa1986, Takeuchi2002}, which depends on the difference between the gas azimuthal velocity and the local Keplerian speed $v_K$, and can be expressed in terms of the pressure gradient through $\eta = -\, (1/2)\,  (h_\textrm{g} / r)^2\, \partiallogdiff{P}$, where $P$ is the pressure. 

In principle, the drift term leads to pebble-sized dust grains moving inward in a smooth disk, which will lead to very small dust disk sizes over time, very low millimeter fluxes, and high spectral indices \citep{Pinilla2012, Pinilla2020}. In contrast, when disks have pressure bumps, particles can be trapped in pressure maxima, which we commonly refer to as dust traps \citep{Weidenschilling1977, Pinilla2012}, while dust diffusivity counterbalances the trapping efficiency, preventing the rings from becoming infinitely narrow.

In our models, the dust grains of different sizes interact between each other through coagulation and fragmentation following the Smoluchowski equation \citep[see][for details]{Birnstiel2010, Stammler2022_Dustpy}, which leads to two distinct regimes in which the local maximum grain size may be either limited by either by fragmentation or drift \citep[see review by][]{Birnstiel2016_Review}. We do not include the effect of back-reaction of dust onto the gas, which can affect disk evolution when there are high concentrations of dust in the disk \citep[e.g.,][]{Garate2020}.

\subsection{Initial conditions and substructure model}

We initialized the disk surface density profile using  a variation of the \citet{Lynden-Bell1974} self-similar solution,
\begin{equation} \label{eq_LBPprofile}
    \Sigmagas(r) = \frac{M_\textrm{disk}}{2\pi r_c^2} \left(\frac{r}{r_c}\right)^{-1} \exp(-r/r_c) \frac{\alpha_0}{\alpha(r)},
\end{equation}
where $M_\textrm{disk}$ and $r_c$ correspond to the initial disk mass and characteristic radius. We included bump-like substructures $\alpha(r)$ in our turbulence profile that lead to self-sustained gaps in the gas surface density profile \citep[as in][]{Stadler2022}. The radial turbulence profile is described by the following expression: 
\begin{equation} \label{eq_alpha_bump}
    \alpha(r) = \alpha_0 \times \left(1 + \sum_{i} A_\textrm{gap} \exp\left(-\frac{\left(r - r_\textrm{gap,i}\right)^2}{2w_\textrm{gap}^2}   \right)\right),
\end{equation}
which accounts for multiple Gaussian substructures of amplitude $A_\textrm{gap}$ and location $r_\textrm{gap,i}$, imposed over a base turbulence parameter of $\alpha_0$. The width of the traps $w_\textrm{gap}$ is defined by setting the Gaussian FWHM to the local scale height value  at the gap location $h_\textrm{g} (r_\textrm{gap})$. As the disk evolves, the viscous evolution will increase the disk characteristic radius, but  keep the gap-like substructures that act as dust traps. The work by \cite{Pinilla2021_b} compared the effect of including static pressure bumps from the beginning of the simulations vs. pressure bumps as assumed in our models. As pressure bumps introduced using Eq.~\ref{eq_alpha_bump} can take more than 1\,Myr to reach the final amplitude, they trap less dust than static traps that introduce early in the simulations. This affects the final observables, such as millimeter fluxes and spectral indices, which are discussed in Sect.~\ref{sec:discussion:dust}.

% Limitations
This approach has the limitation that it does not capture the 3D structure of the dust dynamics, which, for example, can result in more efficient diffusion across the gap when the azimuthal dimension is considered \citep{Weber2018}. Similarly, our approach does not include mechanisms such as photoevaporation or MHD winds, which can contribute to the disk evolution \citep[e.g.,][]{Clarke2001, Fedele2010, Alexander2014, Ribas2015, Tabone2022, pascucci2023}.

We assumed that the disk temperature is set by stellar irradiation and neglected the contribution of other alternative sources, such as viscous heating or external irradiation. Therefore, the temperature profile is given by the following power-law profile:
\begin{equation}
    T(r) = \left(\frac{L_\star}{L_\odot}\right)^{1/4} \left(\frac{r}{\SI{1}{au}}\right)^{-1/2} \SI{263}{K},
\end{equation}
which depends on the stellar luminosity $L_\star$.\par

Finally, the initial dust size distribution follows the interstellar medium distribution \citep{Mathis1977}, with a radially uniform dust-to-gas ratio of $\epsilon_0$, which then evolves through the simulations. The initial dust distribution goes between $0.5-1 $ $\mu$m, and the material density of the dust grains is $\rho_s = 1.6$ g\,cm$^{-3}$, with a fragmentation velocity of $v_\textrm{frag} = 10$ m\,s$^{-1}$ \citep{Gundlach2011, Wada2011, Gundlach2015}, though laboratory studies suggest that this value might be lower \citep{Gundlach2018, Musiolik2019, Steinpilz2019}. Dust grains could still grow to larger sizes even at lower fragmentation velocities, as long as the turbulence near the midplane is also reduced \citep{Pinilla2021_b}.

%%%%%%%%%%%%%%%%%%%%%%%%%%%%%%%%%%%%%%%%%%%%%%%%%%%
%%%%%%%%%%%%%%%%%%%%%%%%%%%%%%%%%%%%%%%%%%%%%%%%%%%
\subsection{Parameter Space}
Our simulations explore the parameter space in the stellar mass $M_\star$, the disk initial mass-to-star ratio ($M_\textrm{disk}/M_\star$), the disk characteristic size $r_c$, and the different gap amplitudes $A_\textrm{gap}$. The ranges covered by these parameters are wide enough to represent the observed properties of our systems. For simplicity, we also refer to the trap amplitudes of Eq.~\ref{eq_alpha_bump} as \sgquote{None} ($A_\textrm{gap} = 0$), \sgquote{Weak} ($A_\textrm{gap} = 1$), or \sgquote{Strong} ($A_\textrm{gap} = 4$). We also explore the parameter space in the $\alpha_0$ turbulence, and initial dust-to-gas ratio $\epsilon_0$.
The values explored are shown in Table \ref{Table_OverviewParameters}. 
It is important to note that these simulations were performed during the time that the AGE-PRO program was receiving data, and for this reason, the initial gas disk mass ($M_\text{gas}$) was not tailored to the particular values that are inferred from these observations \citep[as presented in][]{AGEPRO_V_gasmasses}. Simulations that are tailored to the properties of each disks will be part of a future publication.

\begin{table}
\centering
\caption{Disk model parameters.}
\label{Table_OverviewParameters}
%\begin{center}
\begin{tabular}{l l c}
 \hline \hline
Symbol & Description [Units] & Value \\
 \hline
$M_\star$ & Stellar mass [$M_\odot$] & 0.25, 0.5, 0.75, 1.0 \\
$L_\star$ & Stellar Luminosity [$L_\odot$] & 0.15, 0.25, 0.5, 1.0  \\
$M_\mathrm{disk}$ & Initial disk mass [$M_\star$] & 0.01, 0.05, 0.1  \\
$r_\mathrm{c}$ & Characteristic radius [au] & 15, 30, 60, 120  \\
$A_\mathrm{gap}$ & Trap amplitude & 0, 1, 4\\
$r_\mathrm{gap}$ & Gap locations [au]& 10, 40, 70\\
$\alpha_0$ & Viscosity parameter & $10^{-4}$, $10^{-3}$  \\
$\epsilon_0$ & Initial dust-to-gas ratio & $0.01$, $0.05$ \\
\hline
\end{tabular}
%\end{center}
\vspace{-3.0mm}
\end{table}

The stellar masses range from 0.25\,$M_\odot$ to 1.0\,$M_\odot$ in steps of 0.25, which covers the stellar masses for the AGE-PRO sources. For each stellar mass, we assign the corresponding stellar luminosity (matched in the order shown in Table~\ref{Table_OverviewParameters}), which is used to determine the disk temperature in our dust evolution model. In addition, for each combination of disk mass, we initiate simulations with four different $r_c$, which therefore includes simulations of low-mass stars with very compact disks ($M_\star=0.25\,M_\odot$ and $r_c=15$\,au), as well as solar-type stars with very extended disks ($M_\star=1.0\,M_\odot$ and $r_c=120$\,au). All of the other possible combinations, such as low-mass stars with extended disks, and solar-type stars with compact disks, are also considered. 

The locations of the gaps (when $A_\textrm{gap} > 0$) are fixed at \SI{10}{}, \SI{40}{}, and \SI{70}{au}, provided that $r_\textrm{gap} \leq 2\,r_c$. That means that the disks with $r_c= \SI{15}{au}$ only have one gap at $r_\textrm{gap} = \SI{10}{au}$, the disks with $r_c= \SI{30}{au}$ have two gaps,  while the disks with $r_c= 60, 120\,$au have all three gaps at the same time. We include gaps only inside $2 r_c$, since $83\%$ of the total disk mass is contained within this radius in a disk that follows the same prescription as in \cite{Lynden-Bell1974}, and therefore traps farther outside would only affect a minor fraction of the inward-drifting solids.

The turbulence value associated with the dust dynamics (settling, diffusion, and fragmentation) is typically the same as the base value of the gas turbulence $\alpha_0$. However, for the particular case with low gas turbulence $\alpha_0 = \SI{e-4}{}$ we chose to decouple the turbulence value associated with the dust fragmentation, in order to avoid an overly efficient grain growth regime, which is computationally expensive to trace. For this reason, we impose that the fragmentation turbulence must be at least $\alpha_\textrm{frag} = \SI{5e-4}{}$, while turbulence values associated with radial diffusivity and settling are always set to the value of the gas $\delta_\textrm{rad} = \delta_\textrm{vert} = \alpha_0$.

%%%%%%%%%%%%%%%
%LOCAL ADJUSTMENTS
%%%%%%%%%%%%%%%

\subsection{Numerical grid}

The standard radial grid of the simulations goes from $r_\textrm{in} = \SI{5}{au}$ to $r_\textrm{out} = \SI{500}{au}$ with $n_r = 250$ logarithmically spaced grid cells. This grid is used for most of our simulations, except those with a small characteristic radius $r_c = \SI{15}{au}$, where the grid goes from $\SI{2.5}{au}$ to $r_\textrm{out} = \SI{250}{au}$, to prevent losing a significant fraction of the mass in the inner regions.

The standard mass grid for the grain size distribution ranges from $m_\textrm{min} = \SI{e-12}{g}$ to $m_\textrm{max} = \SI{3e6}{g}$, which, in terms of the grain sizes, corresponds to approximately to $a_\textrm{min} \approx \SI{0.5}{\mu\, m}$ and $a_\textrm{max} \approx \SI{75}{cm}$, with $n_m= 134$ logarithmically spaced grid cells. For the simulations with low turbulence $\alpha_0 = \SI{e-4}{}$, the upper limit of the mass grid is extended to $m_\textrm{max} = \SI{e8}{g}$ (and $n_m = 141$ grid cells).

\begin{figure}
\centering
\includegraphics[width=8cm]{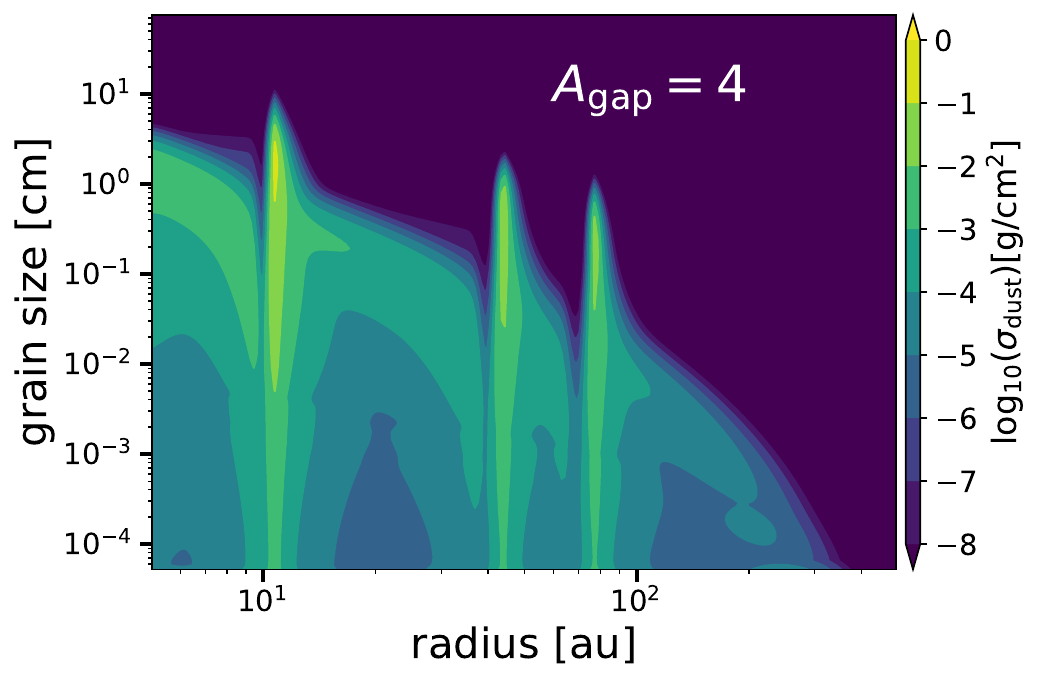}
\includegraphics[width=8cm]{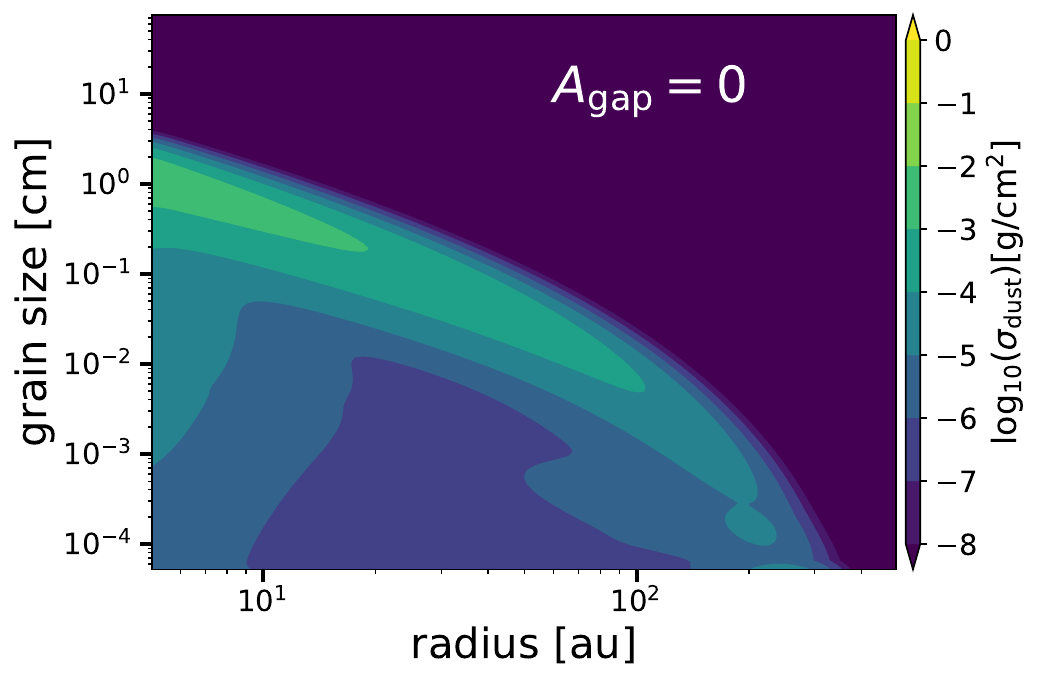} 
\caption{Dust size distribution for a disk with (top) and without (bottom) substructures at $1$ Myr. The simulations correspond to a disk with $M_\star =  1\,{M_\odot}$, $M_\textrm{disk} = 1\,{M_*}$, $r_c = \SI{60}{au}$, $\alpha_0 = \SI{e-3}{}$, and $\epsilon_0 = \SI{0.01}{}$.
 }
 \label{Fig_DustSizeDist}
\end{figure}

The simulations ran from time $t=0$ and up to $t=10$\,Myr.
The key output of the \texttt{DustPy} simulations is time evolution of the  dust surface density, for every dust species in the grain size distribution. We show an example of the dust size distribution in a disk with gaps and without gaps in Figure \ref{Fig_DustSizeDist}, where dust particles are trapped at the outer edges of the gaps.

Considering the different combinations of $\alpha_0$ and $\epsilon_0$ values, we have two groups of simulations with $\alpha_0=10^{-3}$ and different $\epsilon_0$, and one with $\alpha_0=10^{-4}$ and $\epsilon_0=0.01$. Each of these three groups has 144 disks, which result from choosing the different $M_\star$, $M_{\text{disk}}$, $R_{\rm c}$, and $A_{\rm gap}$. The state of each disk is saved at nine different ages (0.1, 0.5, 0.8, 1, 2, 3, 5, 7, and 10\,Myr), thus resulting in 3888 different snapshots of the disks evolution.  \vspace{1cm}

\begin{figure}
\centering
\includegraphics[width=8cm]{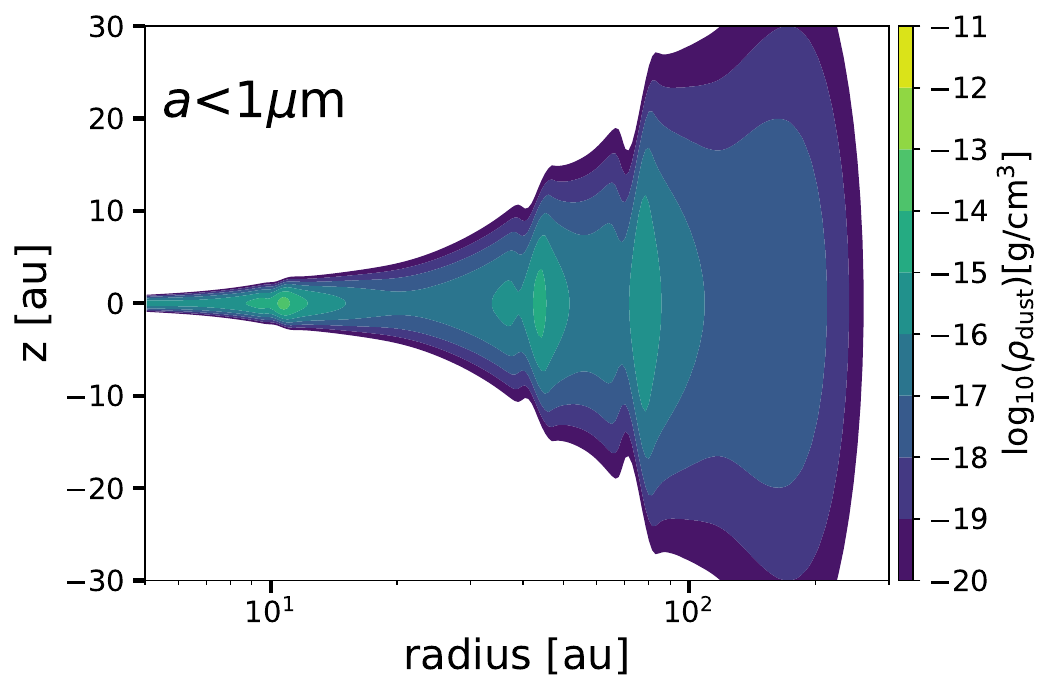}
\includegraphics[width=8cm]{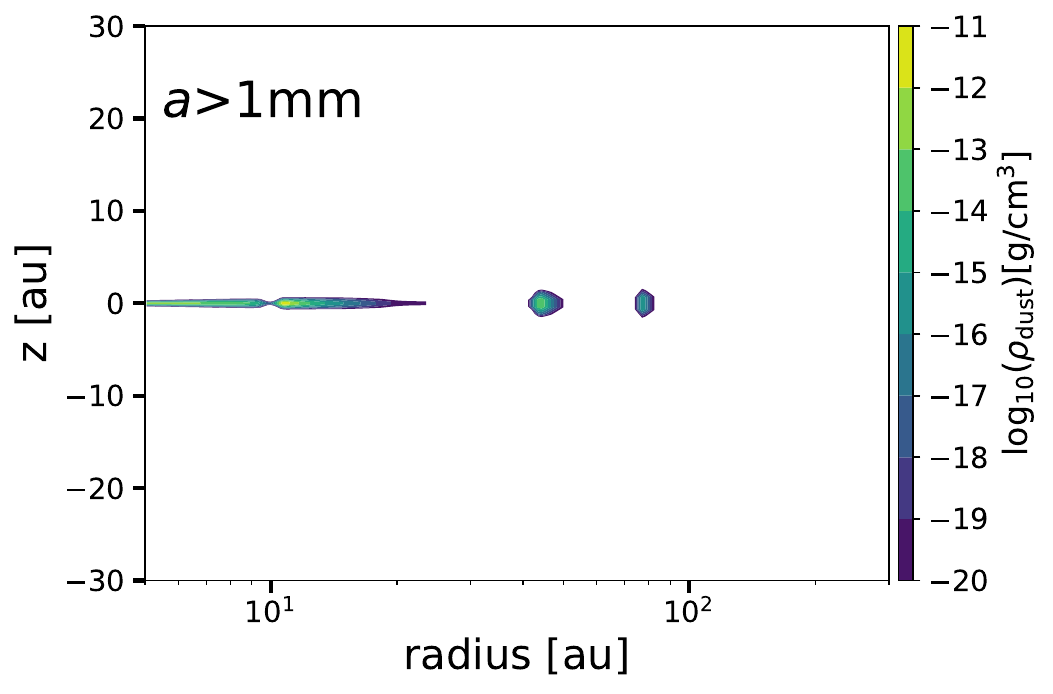}
 \caption{Dust spatial distribution of small (top, $1$\,$\mu$m) and large (bottom $1$\,mm) grains, for a disk with substructures at \SI{1}{Myr} from the same model presented in the top panel of Fig.~\ref{Fig_DustSizeDist}.}
 \label{Fig_DustSpatialDist}
\end{figure}

\subsection{Radiative Transfer}

The millimeter flux of the dust continuum emission is obtained from the simulated dust size distribution with the radiative transfer code RADMC-3D \citep{RADMC3D2012}. 
The main goal of obtaining these fluxes is to compare with the observations of AGE-PRO, and thus we chose to generate images at wavelengths of $\lambda = 1.05\,$mm and $\lambda= 1.3$\,mm, matching the two spectral setups of AGE-PRO. Given the large number of simulations, we select only 5 ages per disk to run the radiative transfer, at $0.5$, $1$, $2$, $5$, and $10$\,Myr, covering all the relevant ages of the star-forming regions (SFR) in the AGE-PRO sample (Ophiuchus, Lupus, Upper Sco, \citet{AGEPRO_I_overview}). This implies a total of 19440 radiative transfer models ($5\times3888$).

% OPACITIES
To obtain the dust opacities, we use the code \texttt{OpTool}\footnote{  \href{https://github.com/cdominik/optool}{github.com/cdominik/optool}}\citep{Dominik2021}, and assume that the grain composition follows the model of \cite{Ricci2010}, where the dust grains are composed of 10\% Silicate \citep{Draine2003}, 20\% Carbon \citep{Zubko1996},  30\% water ice \citep{warren&brandt08},  and 40\% vacuum.\par

%
% GRAIN RESAMPLING
Rather than calculating the opacities for all 134 (or 141) dust grain species, we chose to re-sample our grain size grid (and the corresponding dust surface densities) down to only 30 dust grain species, which still cover all of the key features of the opacity profile from \cite{Ricci2010} at our target wavelength. 
The size grid for the radiative transfer model is defined with three intervals of logarithmically spaced grid cells between $0.5\,\mu$m, $50\,\mu$m, $0.1$\,cm and $20$\,cm, with $5$, $15$, and $10$ grid cells, respectively. Dust sizes that fall outside the grid boundaries are ignored in the re-sampling, with no significant loss of mass.

% VERTICAL EXPANSION OF THE GRID
The 1D dust surface density profiles are converted into 2D volumetric density profiles, in a spherical grid where the radial coordinates are the same as the \texttt{Dustpy} radial grid, and the colatitude coordinate goes from $\theta_{\rm min} = 0$ to $\theta_{\rm max} = \pi$, with $n_\theta = 180$ grid cells. The disk is assumed to be axisymmetric, and we
we also assume that the dust is in vertical hydrostatic equilibrium distributed according to the following Gaussian profile \citep{Fromang2009}:

\begin{equation} \label{eq_vertical_density}
\rho_\textrm{d,i}(r, z) = \frac{\Sigma_\textrm{d,i}(r)}{\sqrt{2\pi} h_\textrm{d,i}(r)} \exp\left(-\frac{z^2}{2 h^2_\textrm{d,i}(r)}\right),
\end{equation}
where $\rho_\textrm{d,i}$ is the volumetric dust density for the dust grain species $i$, $\Sigma_\textrm{d,i}$ is the dust surface density, $h_\textrm{d,i}$ is the dust scale height from \citet{Dubrelle1995}, and $z$ is the vertical distance from the midplane. 
We illustrate the spatial distribution of small and large grains for the case of a disk with substructures in Figure \ref{Fig_DustSpatialDist}, where we observe how larger grains have migrated inward and are settled closer to the midplane.
%Star emission
The central star is assumed to emit as a perfect blackbody, where the total luminosity is the corresponding value of $L_\star$, and the stellar temperature is obtained from the stellar tracks \citep{dotter2008}.\par

%RADMC SETUP
The radiative transfer calculation was performed in two steps. First, we ran a Monte Carlo simulation to obtain the dust temperature for every grain species, using $n_\textrm{phot} = 10^7$ photon packages. In a second step, we obtained the simulated images at the target wavelengths, with $n_\textrm{phot} = 2\cdot10^5$ photon packages to account for the contribution of isotropic scattering. 
We ignore more complex scattering processes due to the high computational cost for covering our complete parameter space, and we verified the flux variations were well below 10\% of the total flux. 

For additional testing, we also further processed the radiative transfer images into synthetic observations to simulate the sensitivity and angular resolution of the AGE-PRO observational setup. In the main text, all the comparisons are done between the radiative transfer models and the properties of the AGE-PRO disks as recovered from the observations, and additional discussion on the influence of the beam convolution when comparing disk radii is provided in the Appendix.

\begin{figure*}
\centering
\includegraphics[width=1.0\textwidth]{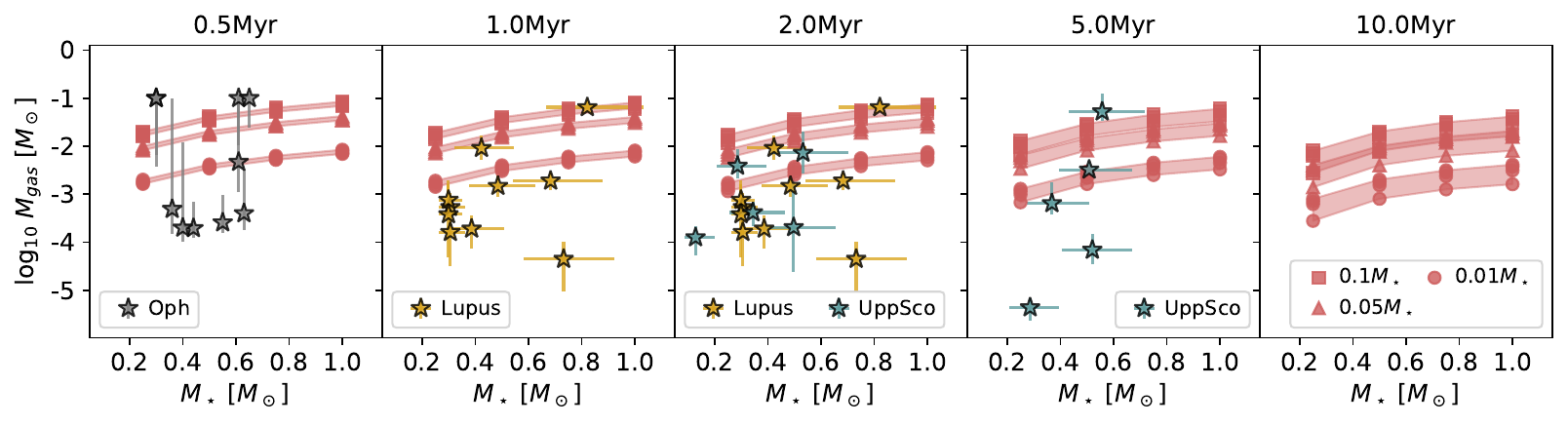}
 \caption{ Evolution of disk gas mass ($M_{\rm gas}$) as a function of stellar mass ($M_\star$) for different times of evolution, as shown at the top of each panel. 
 The shaded regions enclose the disks with the same initial disk mass relative to their host star, and the markers show the maximum, median, and minimum value. 
 The spread in $M_{\rm gas}$ for a fixed stellar mass comes from the different initial disk characteristic size $r_c$ within the same group of initial disk mass, and its time evolution. 
 The plot features only the simulations without traps ($A_\textrm{gap} = 0$), standard turbulence $\alpha_0 = 10^{-3}$, and interestellar medium dust-to-gas ratio of $\epsilon_0 = \SI{0.01}{}$. The disks with traps, and those from the alternative simulation setups show a comparable behavior. 
 For comparison, the gray square markers show the estimated values for $M_{\rm gas}$ for the disks in Oph (for the 0.5 Myr snapshot), Lupus (for the 1 and 2 Myr snapshots), and Upper Sco (for the 2 and 5 Myr snapshots) as obtained in \citet{AGEPRO_V_gasmasses}.}
 \label{fig:mstar_mgas_age}
\end{figure*}

%%%%%%%%%%%%%%%%%%%%%%%%%%%%%%%%%%%%%%%%%%%%%%
\section{Results} \label{section_Results}
%%%%%%%%%%%%%%%%%%%%%%%%%%%%%%%%%%%%%%%%%%%%%%

In the following Section, the comparison between the AGE-PRO sources and our simulations is done by groups of different ages. The Ophiuchus sources are compared to our simulations at 0.5\,Myr. The Lupus sources are compared to the simulations at 1\,Myr and 2\,Myr. The Upper Sco sources are divided in two groups: For stellar hosts younger than 3\,Myr, we compared them to our simulations at 2\,Myr, while for the stellar hosts older than 3\,Myr, we compared them to the simulations at 5\,Myr. The decision to split at these ages comes from the recalculation of stellar properties for the sources within AGE-PRO, which resulted in two age clusters in the Upper Sco sample, one of which is more consistent with the ages of the Lupus disks, while another is consistent with an older age. The full discussion about the estimation of $M_\star$ and age is given in \citet{AGEPRO_I_overview}, and we refer the reader to that work for a detailed description. 
We compared the observational results of AGE-PRO to the simulations with standard values of $\alpha=10^{-3}$ and dust-to-gas ratio of $0.01$. While the specific properties of the disks can change slightly for $\alpha=10^{-4}$ or initial dust-to-gas ratio of $0.05$, the qualitative evolution and spread of properties is the same as with the standard values. 

The following subsections focus on comparing the results of the simulations to those of the observed disks in AGE-PRO. For each comparison, we present the results of the simulations in groups of disks with the same initial condition, such as their initial disk mass as a fraction of stellar mass ($M_{\rm disk}/M_\star$), or by their initial disk characteristic size $r_c$.

\subsection{Gas mass evolution}

In our simulations, the evolution of $M_{\rm gas}$ is only driven by viscous evolution. The presence of dust traps does not produce a significant difference in the evolution of $M_{\rm gas}$, and thus, we only compare the disks masses measured from AGE-PRO to those of our disks without traps ($A_\textrm{gap} = 0$), as shown in Figure \ref{fig:mstar_mgas_age}. When comparing against $M_{\rm gas}$ derived from \citet{AGEPRO_V_gasmasses}, we find that the latter have a larger dispersion in their values compared to those found in our simulations. In fact, the measured $M_{\rm gas}$ from observations spans $\sim3$ orders of magnitude in Ophiuchus, while our initial disk conditions ($t=0$\,Myr) only span one order of magnitude for each stellar mass ($M_{\rm gas}$ from $0.01\,M_\star$ to $0.1\,M_\star$). The lowest disk masses in Ophiuchus have 2 orders of magnitude less gas mass than those of comparable $M_\star$ in our simulations, and only about half of the targets in Lupus and Upper\,Sco are in agreement to the measured $M_{\rm gas}$ of the models. This mismatch between the trends of $M_{\rm gas}$ evolution prevents further comparisons between $M_{\rm gas}$ and other disk or stellar parameters, such as the trends analyzed in \citet{AGEPRO_V_gasmasses}. This topic is further discussed in detail in Sect.~\ref{sect:discussion}, as well as in \citet{AGEPRO_VII_population} and \citet{AGEPRO_VIII_ext_photoevap}.

\subsection{Dust mass comparison}\label{sec:comp_dustmass}

\begin{figure*}
  \centering
  \includegraphics[width=1.0\textwidth]{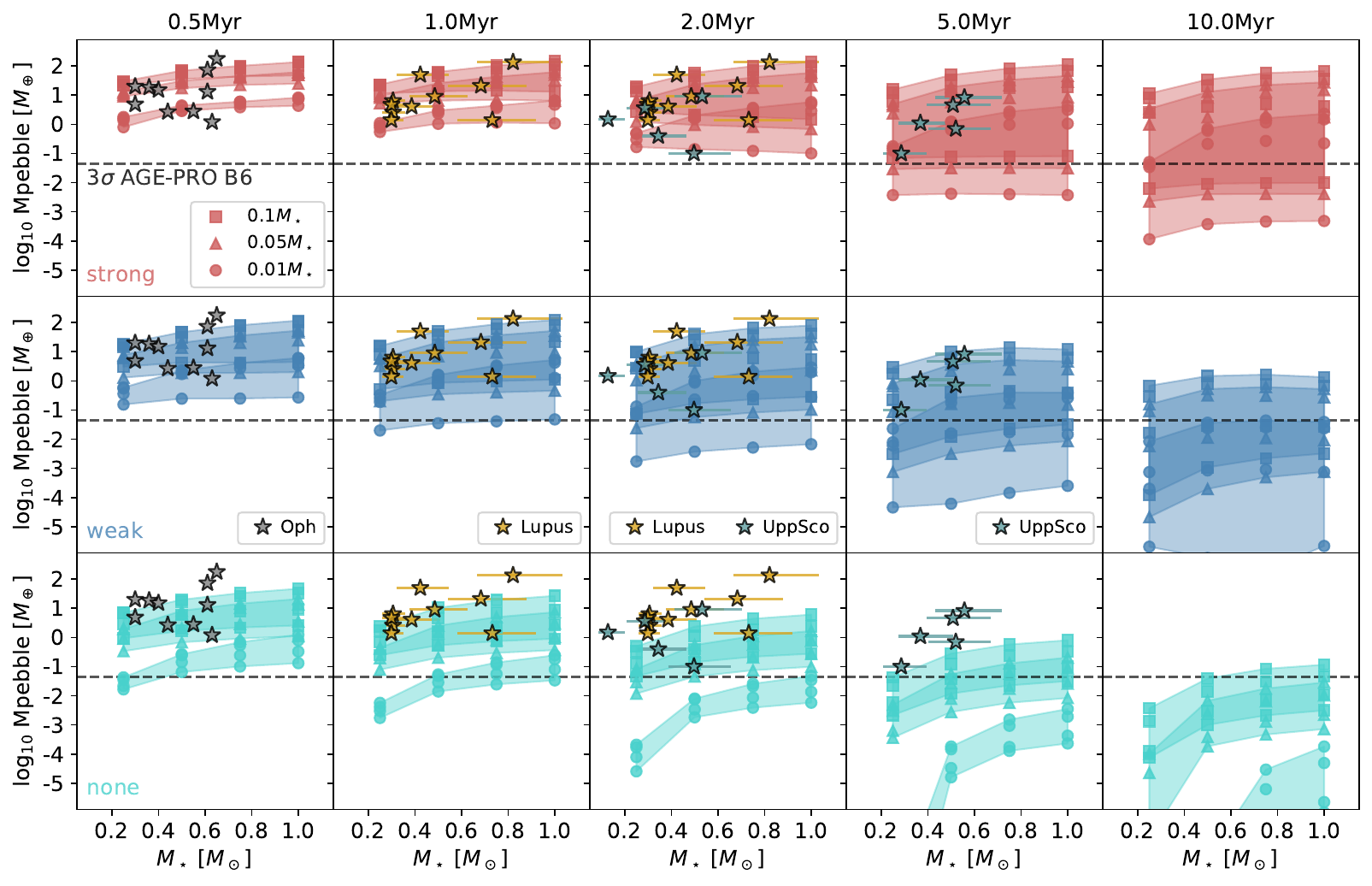}
  \caption{Pebble mass vs. stellar mass at different ages and trap amplitudes (strong, weak, and none from top to bottom, respectively) from simulations, considering pebbles as grain sizes between $0.1$\,mm and $10$\,cm. The shaded regions enclose the simulated disks of different initial disk masses. 
  For comparison, the star markers show the estimated values for $M_{\rm{dust}}$ for the disks in Ophiuchus (compared to 0.5 Myr), Lupus (compared to 1 and 2\,Myr) and Upper Sco (compared to 3 or 5\,Myr, depending on the age of the sources). The dashed horizontal line shows the approximated $3\sigma$ sensitivity of AGE-PRO. }
  \label{fig:mstar_mpebble}
\end{figure*}

\begin{figure*}
  \centering
  \includegraphics[width=1.0\textwidth]{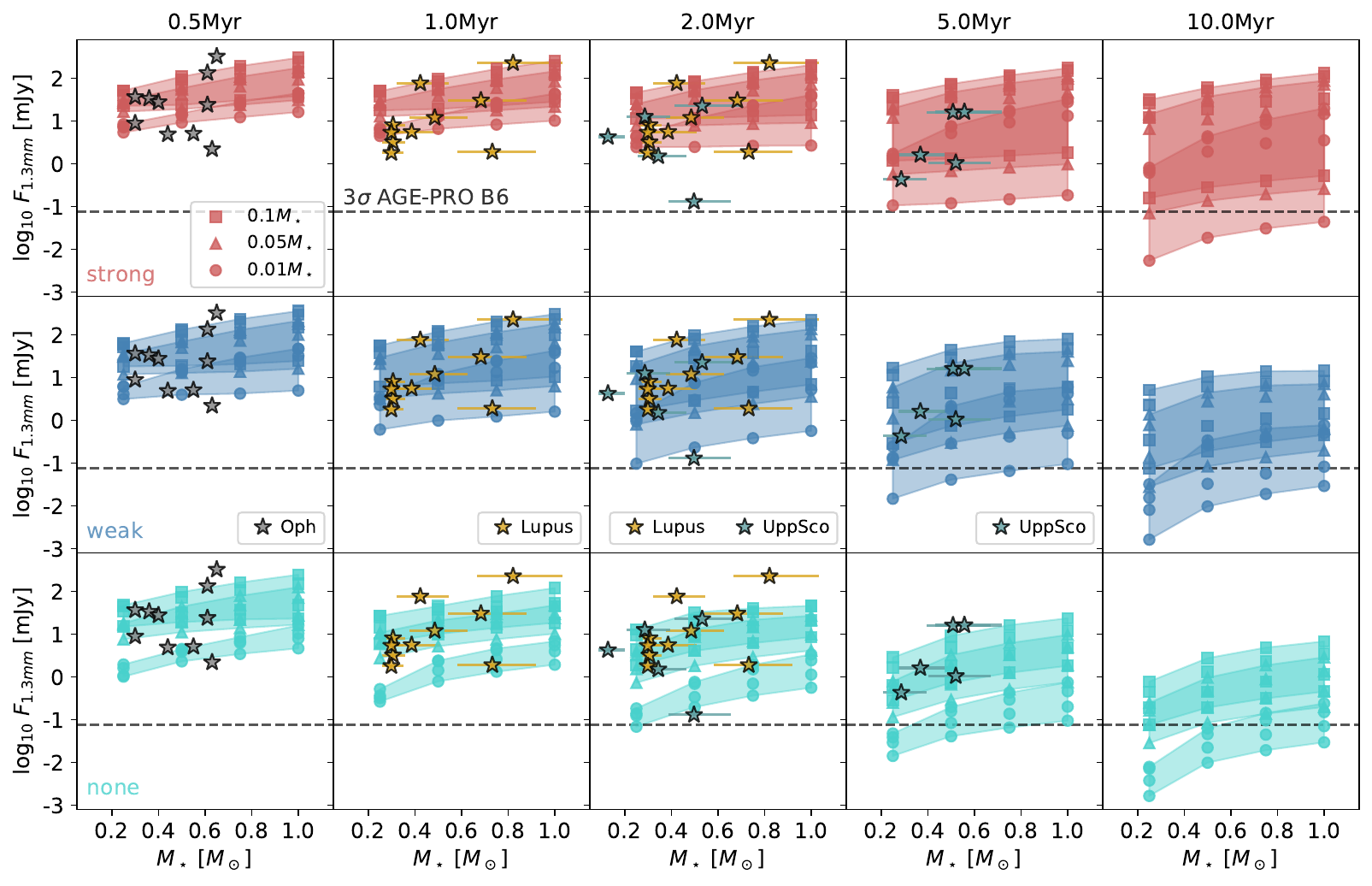}
  \caption{Same as Figure \ref{fig:mstar_mpebble}, but with dust continuum fluxes at 1.3\,mm ($230$\,GHz) vs. the stellar mass. }
  \label{fig:mstar_1.3mmfluxes}
\end{figure*}

Despite the difference in distribution between modeled and observed gas masses, we find consistent values for the 1.3\,mm dust continuum brightness among the AGE-PRO targets and our models, which were estimated from the radiative transfer models. In observations, the dust mass content in pebbles can be roughly approximated from the disk dust continuum emission using \citep{hildebrand1983}:
\begin{equation}
    M_{dust} \, = \, \frac{d^2\,F_\nu}{\kappa_\nu\,B_\nu(T)} \text{,}
    \label{eq:mdust}
\end{equation}
\noindent where it is assumed that the emission is optically thin, where is $d$ the distance to each star, measured as the inverse of the parallax from Gaia DR3 \citep{gaia2021edr3}, $\nu$ is the frequency of the AGE-PRO Band 6 observation (1.3\,mm, $230$\,GHz), $B_\nu$ is the Planck function, and $\kappa_\nu$ is the mass absorption coefficient, which is typically assumed to be $\kappa_\nu=2.3\cdot(\nu / 230\,\text{GHz})^{0.4}\,\text{cm}^2\text{g}^{-1}$, as in \cite{andrews2013}. For the temperature, a constant $T=20\,$K is used for consistency with the literature \citep[e.g.,][]{Ansdell2016}. The dust content of the AGE-PRO disks was calculated directly from the measured millimeter fluxes, and they are presented in \citet{AGEPRO_I_overview}, \citet{AGEPRO_II_Ophiuchus}, \citet{AGEPRO_III_Lupus}, and \citet{AGEPRO_IV_UpperSco}. 
The pebble mass in our simulations ($M_{\rm pebble}$) is measured by adding the mass content in particles of sizes ranging from $0.1$\,mm to $10$\,cm, which are the grains that should contribute more efficiently to the emission in the millimeter continuum in the Bands 6 and 7 \citep{Ricci2010}. These results are presented in Fig.~\ref{fig:mstar_mpebble}. 

The simulations with weak or strong traps ($A_\textrm{\rm gap} = 1 - 4$) tend to match the observed dust masses closely over time, although there is a wide spread in $M_{\rm pebble}$ from the simulations, given by the different values of initial disk mass, disk size, and amplitude of dust trapping. 
For the older sources in Upper Sco, the disks are more consistent with the simulations with traps and larger initial disk mass, of at least $0.05\,M_\star$ (see panels showing the 5\,Myr outputs). Overall, disk with strong dust traps and high initial disk masses are able to retain their dust mass over their lifetime evolution, remaining massive in dust even at the 10\,Myr of age. 

In contrast to the strong traps, $M_{\rm pebble}$ in the simulations tends to drop quickly if no traps are present, reaching values that are several orders of magnitude below $0.1 \, M_\oplus$ even by $2$\,Myr for the lowest initial disk masses. This trend also holds at higher initial dust-to-gas ratios of $\epsilon_0 = 0.05$, and at a lower alpha value of $\alpha=10^{-4}$. Thus, the absence of traps results in the loss of dust mass for a wide range of initial conditions and disk viscosity, and these models are unable to explain observations. 

An interesting trend from the models is that the $M_{\rm pebble}$ decrease over time is more significant for lower $M_\star$ in the absence of traps, as previously demonstrated by \cite{Pinilla2013} using a smaller parameter space. However, this trend is not seen in the AGE-PRO observations, potentially due to the presence of unresolved dust traps and the limited sample size.

\subsection{Comparison with observations: Dust continuum emission}

The assumption of optically thin emission and constant dust temperature might oversimplify the conditions of the dust continuum emission in the AGE-PRO sample, especially considering that unresolved dust substructures could become more optically thick, the temperature might not be constant, and the opacity law might be more complex than our simple assumption. 
Thus, comparing the observed flux of the AGE-PRO sources to the radiative transfer model of the simulations is a more robust approach than comparing $M_{\rm pebble}$ to the estimated $M_{\rm dust}$. 

In Figure \ref{fig:mstar_1.3mmfluxes}, the total flux of our simulations at 1.3\,mm is compared to the measurements of AGE-PRO at the same wavelength. The overall conclusion from comparing dust masses remains the same when comparing the fluxes: disks with strong and weak traps are more representative of the evolutionary trend observed across the three regions. However, we also find that the total brightness of some disks could be explained by disks without any dust trapping. The key difference with the dust mass comparison from Sect.~\ref{sec:comp_dustmass} comes from the contribution to the total brightness from small grains ($<0.1\,$mm) in the opacities from \citet{Ricci2010}. See Sect.~\ref{sect:discussion} for a more detailed discussion on this topic.

The 1.3\,mm flux of the three faintest sources in Lupus sample (Lupus 5, 7, 9) is barely consistent with the strong trapping dust models at 2\,Myr of age, and they are not consistent with the 1\,Myr strong trap disks. On the opposite side, the brightest sources in Lupus (Lupus 2, 10) cannot be explained in a model without dust traps at either 1 or 2\,Myr of age. Similarly, the brightest sources of the Upper Sco sample (Upper Sco 1, 8, 9, 10) cannot be explained by models without dust traps if they are older than 5\,Myr. 

\begin{figure*}
\centering
\includegraphics[width=1.0\textwidth]{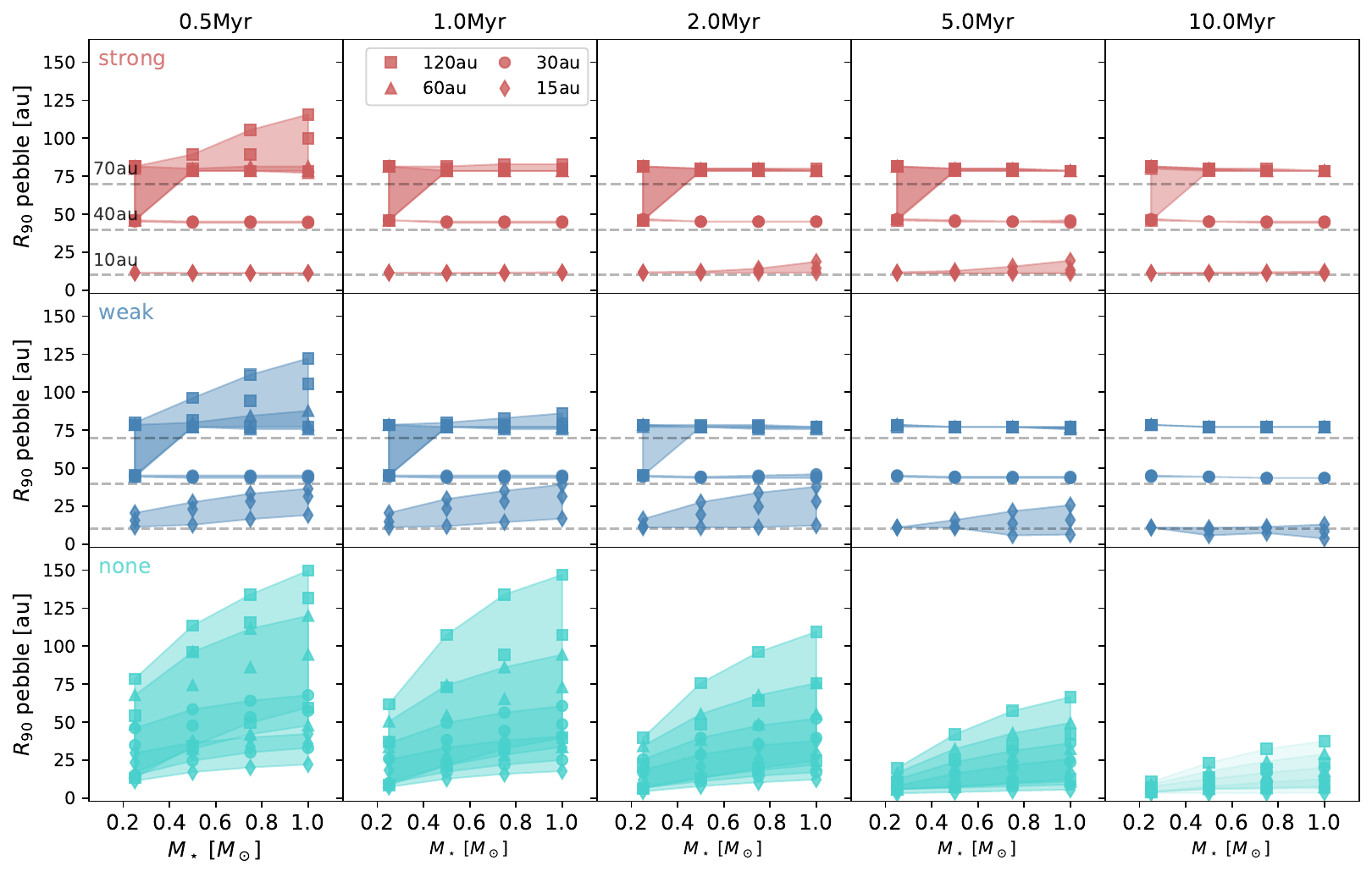}
 \caption{Radius enclosing $90\%$ of the pebble mass vs. stellar mass. The shaded regions enclose the simulated disks of different initial disk critical radius, as shown by their different symbols. 
 The dashed lines show the location of the dust traps in the simulations. 
 }
 \label{fig:mstar_r90pebble}
\end{figure*}

\begin{figure*}
\centering
\includegraphics[width=1.0\textwidth]{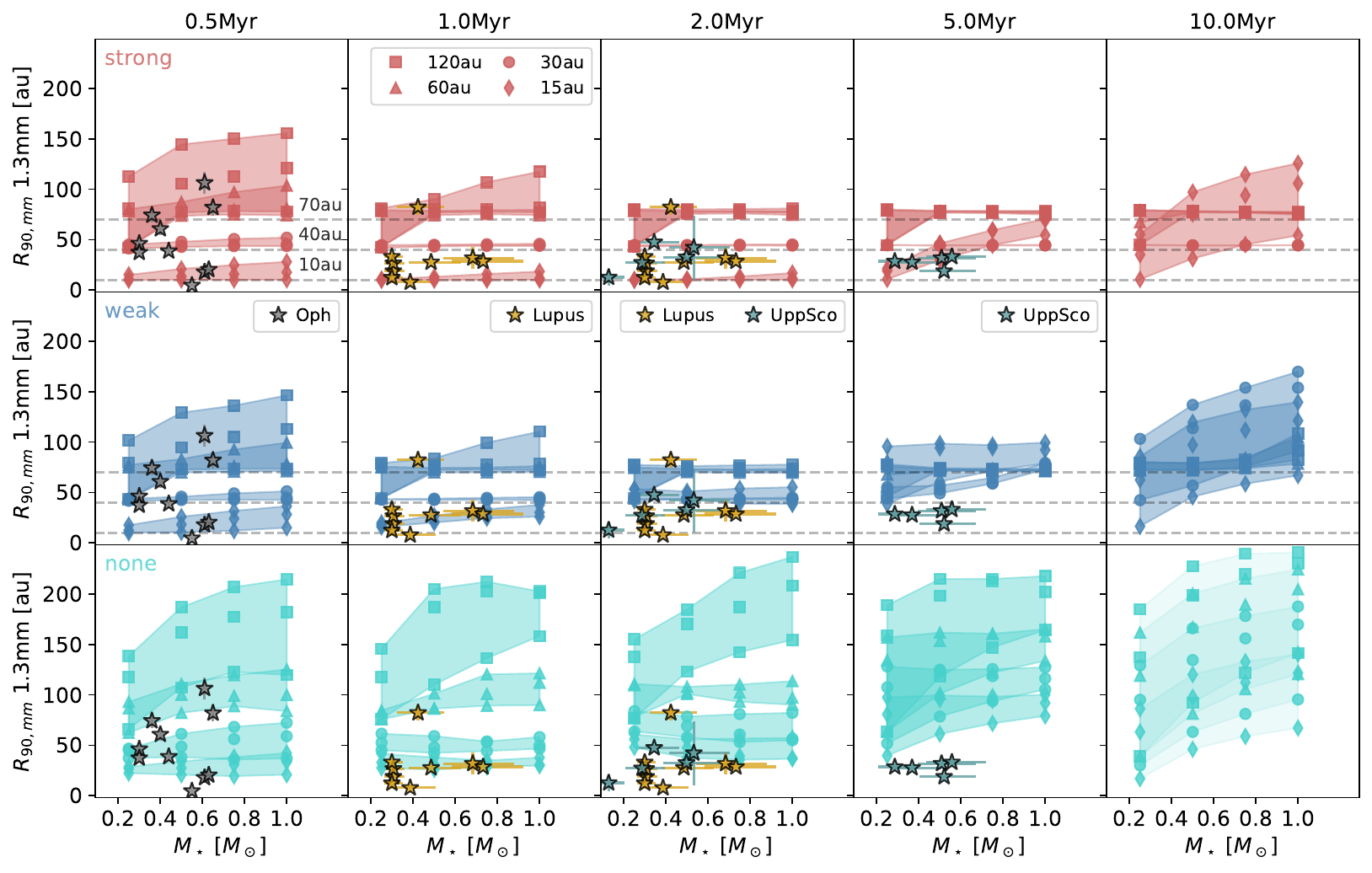}
 \caption{Radius enclosing $90\%$ of the millimeter flux continuum as a function of the stellar mass. The symbols and colors regions are the same as in Figure \ref{fig:mstar_r90pebble}. The star markers show the measured $R_{90\%}$ of the AGE-PRO sources, with the exception of V1094\,Sco (Lupus 10). 
 The dashed lines show the location of the dust traps in the simulations. }
 \label{fig:mstar_rmod90_b6}
\end{figure*}

As in the case of the comparison with the dust mass in Sect.~\ref{sec:comp_dustmass}, we find that the fluxes observed in the AGE-PRO sample are more consistent with the simulations with strong and weak dust traps for ages later than $1$\,Myr. Differently from the comparison with pebble mass, where the dust trapping amplitude could be distinguished even at the early stages at 0.5\,Myr, their dust continuum brightness is consistent with any of the dust trapping scenarios. Therefore, according to our models,  disentangling the potential trapping efficiency in planet-forming disks purely from continuum flux measurements its only possible when looking at the evolution over time, as young disks ($<1\,$Myr) have similar fluxes in the models independent of the presence and amplitude of pressure bumps.

\subsection{Disk size in pebble mass and dust continuum emission}

We estimate the size of the disks in dust content by measuring the radius at which the 90\% of the $M_{\text{pebble}}$ is contained, and we refer to this value as $R_{90\%\text{,pebble}}$, as shown in Figure \ref{fig:mstar_r90pebble}. Because it is derived from $M_{\text{pebble}}$, this $R_{90\%\text{,pebble}}$ only considers pebbles larger than 0.1\,mm. In the disks with strong and weak traps, the location of the radius containing the $R_{90\%\text{,pebble}}$ is completely determined by the location of the outer dust trap, even at early ages of 1\,Myr, and also for the late ages of 10\,Myr. In the case of disks with no dust trap, the $R_{90\%\text{,pebble}}$ decreases as a function of age as the larger pebbles drift unimpeded toward the center of the disk. We further discuss this result in comparison with the literature in Sect.~\ref{sect:discussion}.

From an observational perspective, calculating the $R_{90\%\text{,pebble}}$ is challenging due to the poorly constrained distribution of optical depth, opacity, and temperature. Thus, an alternative measure for the disk size is the radius containing a certain fraction of the brightness, typically chosen to be $90\%$ \citep[e.g.,][]{rosotti2019, hendler2020, kurtovic2021, sanchis2021, long2022}. Further details about these measurements in the AGE-PRO sample are described in \citet{AGEPRO_X_dust_disks} and \citet{AGEPRO_XI_gas_disk_sizes}. We compared these radii containing the $90\%$ of the emission ($R_{90\%,mm}$) from the observations of AGE-PRO to that of our radiative transfer images at 1.3\,mm, as shown in Fig.~\ref{fig:mstar_rmod90_b6} as a function of $M_\star$ and age.

At the early age of 0.5\,Myr, the distribution of $R_{{90\%},mm}$ is similar regardless of the strength of trapping, with the $R_{{90\%},mm}$ only slightly evolving from the initial disk conditions. 
In the case of the disks with traps, at an age of 1\,Myr, the material in the dust traps represents the major contribution to the disk continuum brightness. Thus, $R_{{90\%},mm}$ also traces the location of the outermost trap, both for weak and strong traps.

The role of dust traps in determining $R_{{90\%},mm}$ changes as the radial drift starts depleting the disk of pebbles. At 2\,Myr for the disks with weak traps, and at 5\,Myr for the disks with strong traps, the most compact disks (with a starting critical radius of 15\,au) show an increase in their $R_{\rm {90\%},mm}$, as the relative contribution to the disk brightness reverses from the dust in the traps to the micron-sized grains in the outer regions of the disk. By the age of 10\,Myr, disks with very different initial conditions show similar $R_{\rm {90\%},mm}$, even for groups of disks with the same trap strength. However, these disks can still be distinguished based on their millimeter brightness.

In the case of disks with no dust traps, the $R_{\rm {90\%},mm}$ tends to increase as a function of age, as the micron-sized grains in the outer disk become the dominant contributing source to the millimeter emission when the larger grains are depleted by migrating inward in the disk. This is the opposite as the evolution of $R_{\rm {90\%},pebble}$. 
Similarly to the simulations with dust traps, the disks with smaller initial $r_c$ show a faster increase as a function of age, as solids are depleted at earlier stages. The disks in Upper Sco that are older than 3\,Myr can only be reproduced by the simulations where the outermost dust trap was located closer to the star ($<20$\,au), and they are more consistent with strong trap models.

\subsection{Spectral index}

\begin{figure*}
  \centering
  \includegraphics[width=1.0\textwidth]{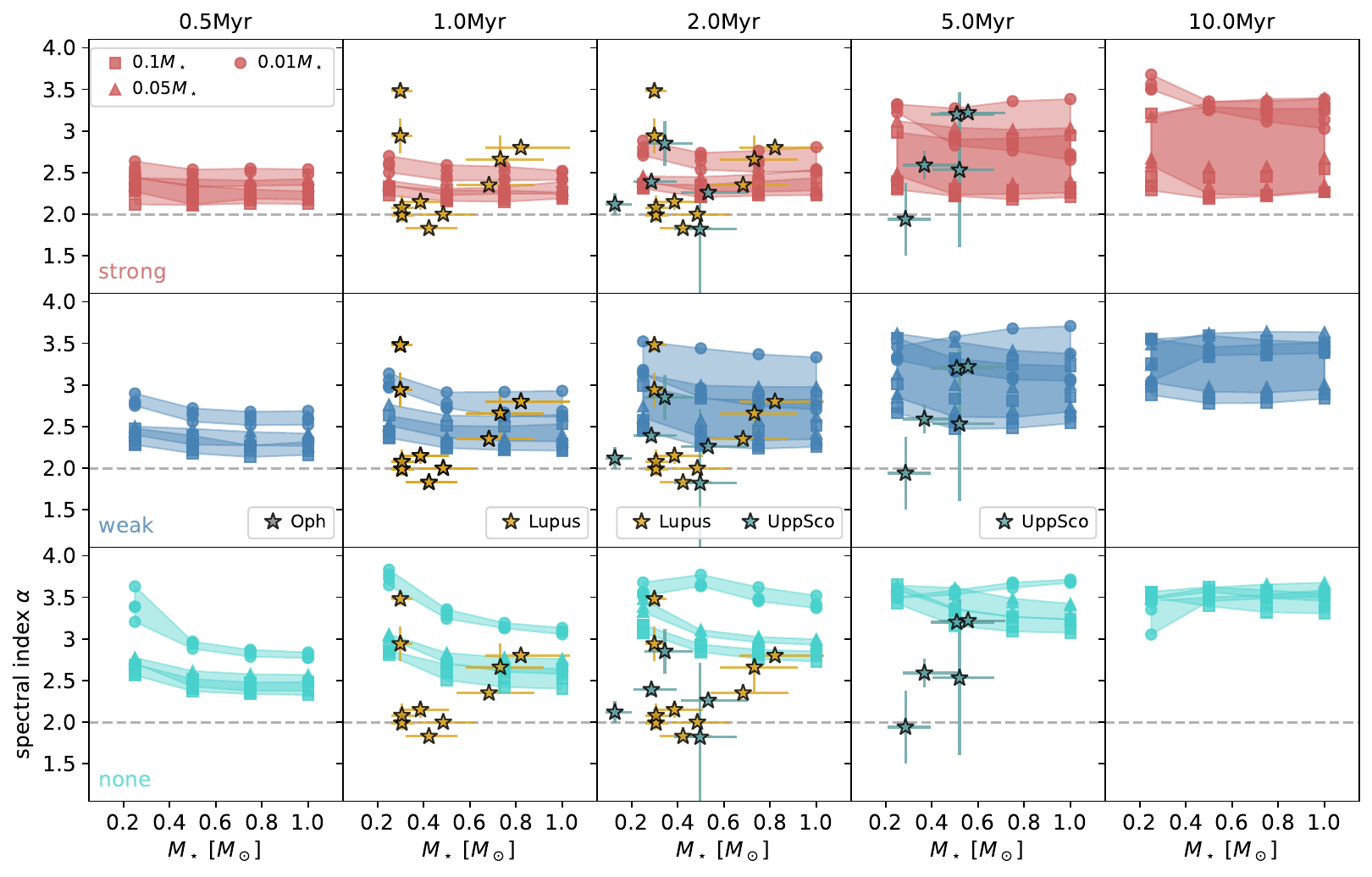}
  \caption{Spectral index between 230GHz and 283GHz (1.3\,mm and 1.06\,mm) vs. stellar mass, representative of the wavelengths covered by the AGE-PRO ALMA Band 6 and Band 7 observations. The colored regions enclose the minimum and maximum spectral indexes for each initial disk mass group. The squares show the spectral index as measured from the AGE-PRO observations for the Lupus and Upper Sco SFR. }
  \label{fig:mstar_specidx_mod}
\end{figure*}

\begin{figure*}
  \centering
  \includegraphics[width=1.0\textwidth]{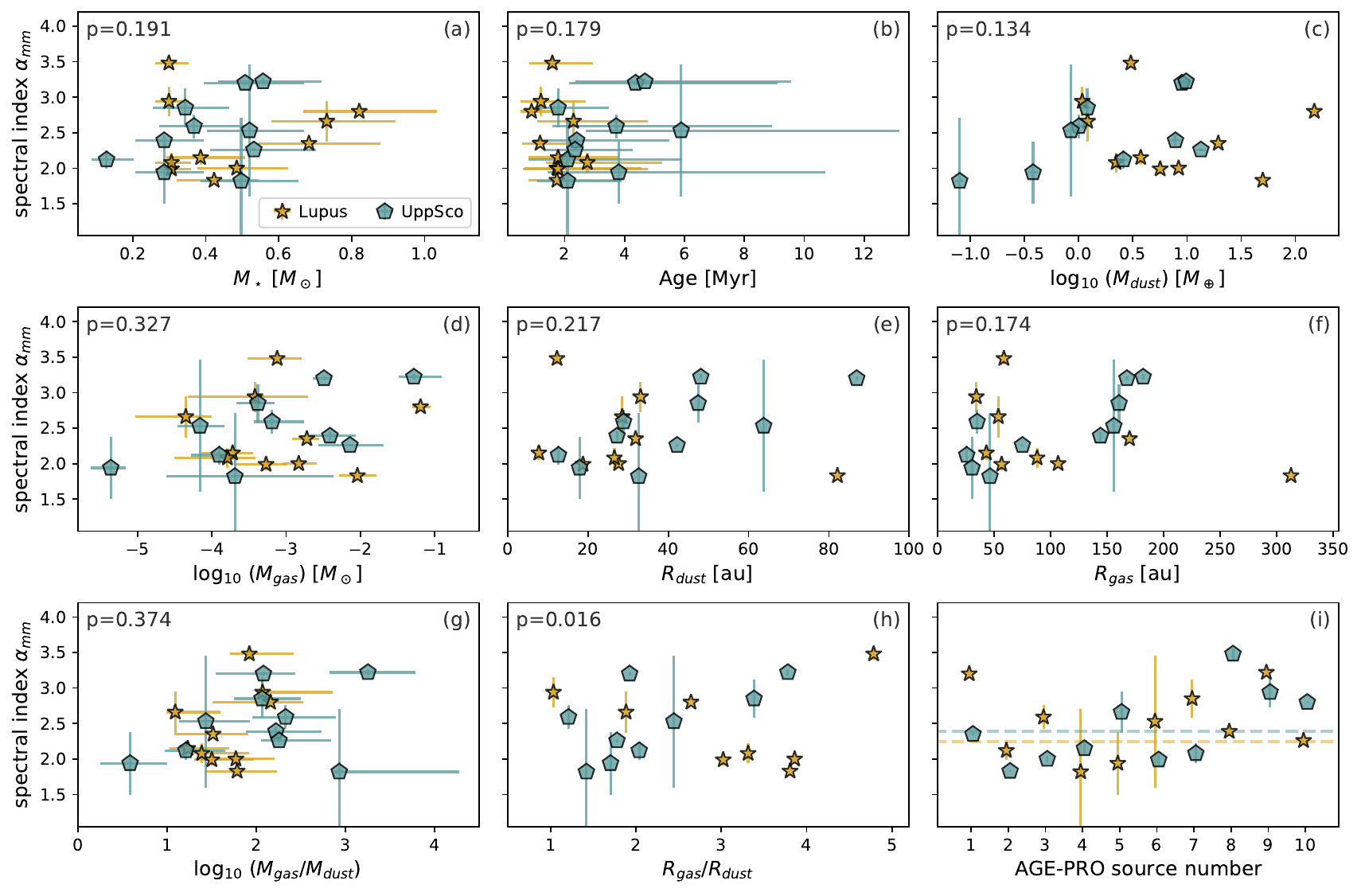}
  \caption{Observed spectral index between 230GHz and 283GHz (1.3\,mm and 1.06\,mm), compared to other observational properties. The dashed lines in panel (i) show the median spectral index for each SFR. }
  \label{fig:mstar_specidx_gallery}
\end{figure*}

The dual-band setup of the AGE-PRO observations (Band 6 and Band 7 at 1.3-1.05\,mm) allows for comparison of the spatially integrated spectral index $\alpha_{\rm{mm}}$ with our radiative transfer images. We measure the spectral index from the total flux of the disks,  assuming Rayleigh-Jeans limit and using the relation: 

\begin{equation}
    \alpha_{\rm{mm}} \, = \,  \frac{ \log_{10} F_{\text{B7}} - \log_{10} F_{\text{B6}}}{\log_{10} v_{\text{B7}} - \log_{10} v_{\text{B6}}} \text{,}
\label{eq:spectral_index}
\end{equation}

\noindent where $F$ is the flux in each ALMA band, and $\nu$ is the representative frequency for that flux. Values of $\alpha_{mm}\approx2$ are expected for disks that are optically thick at the observed wavelengths, while higher values are typically associated with more optically thin emission \citep[see ][]{tazzari2021}. The results are shown in Fig.~\ref{fig:mstar_specidx_mod}.

Interestingly, a few disks in the Lupus region have $\alpha_{\rm{mm}}\approx2$ or below, which are not matched by any of our simulations. These very low values are common in the Lupus SFR \citep{tazzari2021}, and a more detailed description of $\alpha_{\rm{mm}}$  with a wider wavelength coverage might be needed to understand the mechanisms driving these low values. Among the possible scenarios, free-free emission coming from the disk center could contribute to lower $\alpha$ when the total flux is considered \citep[e.g.,][]{garufi2025}, especially in compact disks, and also self-scattering of millimeter emission in optically thick disks \citep{liu2019}. 
Thus, multiwavelength observations at higher angular resolution are needed to recover $\alpha_{\rm{mm}}$ as a function of radii \citep[e.g.,][]{Huang2018, long2020, Pinilla2021, doi2023, das2024, sierra2024}. The remaining disks in Lupus seem consistent with either weak or strong traps, as their the spectral index cannot be reproduced by disks with no traps. 

A large dispersion in the values of spectral indices is also observed in Upper Sco, ranging between 1.9-3.2, which is probably tracing the diversity of locations and strength of dust traps in these disks. Almost all of the values for spectral index in this region are consistent with strong and weak dust traps, but a few disks with the highest spectral index could be consistent with weak traps. Even though some Upper Sco disks have a matching spectral index with simulations with no traps, those disks are not consistent in the other properties (flux or radii) to those simulations, which contributes to rule-out that scenario. Our results are also consistent with those from \cite{Stadler2022} and \cite{delussu2024}, who also demonstrated that dust traps are needed to explain the spectral index of large disk populations. 

The observed spectral indices of Lupus and Upper Sco are also shown in Fig.~\ref{fig:mstar_specidx_gallery} in comparison with multiple properties of those systems. These comparisons show no clear trend with disk gas or dust mass ($M_{\rm gas}$, $M_{\rm dust}$), neither with disk size ($R_{\rm gas}$, $R_{\rm dust}$). A tentative trend, however, is seen between the spectral index and the gas-to-dust mass ratio of the disks (panel g in Fig.~\ref{fig:mstar_specidx_gallery}).  In the Upper Sco SFR, there is a potential relation between larger $R_{\rm gas}$ and higher spectral index, but a larger disk sample is needed to test it.  Similarly, higher angular resolution observations would allow us to explore the relation between spectral index and dust emission morphology, including the location and shape of dust traps that could be driving the large range of total flux measurements.

\section{Discussion} \label{sect:discussion}

\subsection{The evolution of $M_{\rm gas}$}\label{sec:discussion:mgas}

The $M_{\rm gas}$ of the AGE-PRO sample from \citet{AGEPRO_V_gasmasses} are distributed over a wider range of masses compared to our models, which only include the pure viscous evolution scenario, as presented in Fig.~\ref{fig:mstar_mgas_age}. The dispersion of one order of magnitude in the initial mass of our disks, which were set to be initially 1\%, 5\%, and 10\% of the stellar host mass $M_\star$, is not enough to cover the three orders of magnitude of difference that is observed in Ophiuchus (see Fig.~\ref{fig:mstar_mgas_age}), which later increases to be about four orders of magnitude for Lupus and Upper Sco. Within the uncertainties, less than half of the disks are consistent with those of our simulations, and the disks with the lowest $M_{\rm gas}$ do not even coincide with the simulations with initial $M_{\rm gas} = 0.01M_\star$. 

The slow decline of $M_{\rm gas}$ in our simulations is a result of the long viscous timescales in our disks, which are simulated with a constant $\alpha$ of $10^{-3}$ or $10^{-4}$ depending on the setup. While a value greater than $10^{-3}$ value for $\alpha$ would accelerate the disk dispersal by viscous accretion, various indirect observational results suggest that, at a disk population level, the disk turbulence typically has low values of $\alpha \leq 3\cdot10^{-3}$ \citep[see review by][]{rosotti2023}. This is suggested by the absence of evidence for viscous spreading \citep[e.g.,][]{long2022}, the large spread in the $\dot{M}/M_\star$ relation \citep[e.g., ][]{lodato2017}, or the low scale height at millimeter wavelengths measured from edge-on disks \citep{villenave2020, villenave2022}. 

Another caveat when considering viscous timescales in our simulations is that our models have a constant $\alpha$ not only in radii but also in time. A time evolution in the value of $\alpha$ is expected as the disks and their environment evolve, thus changing the distribution of temperature, density, and ionization \citep[e.g.,][]{Delage2022, delage2023}. Very short viscous timescales for ages $<1.0\,$Myr could contribute to removing disk masses, with $\alpha$ values decreasing for later disk stages, explaining the small variation in disk masses between Lupus and Upper Sco, the low $M_{\rm gas}$ for some Ophiuchus disks, and also the large variation between Ophiuchus and Lupus. 

While the star formation process can potentially result in a large diversity of initial conditions, population synthesis models such as \citet{Bate2018} have shown that the initial disk mass is typically in the range between $M_\textrm{disk}/M_\star = 0.1$ to $1$ for ages $<10^5$\,yr. Since it seems unrealistic to expect the initial disk-star-mass ratio to be $<1$\% (especially considering that the sample selection of AGE-PRO required a previous $^{12}$CO detection), the mismatch between our simulations and the AGE-PRO observations suggests that viscous evolution with globally low-viscosity scenarios ($\alpha=10^{-3}\sim10^{-4}$) is not the only responsible for transporting angular momentum and/or removing gas from the disk over its lifetime. Instead, higher-viscosity scenarios, or other processes in addition to viscous evolution, must also be contributing, such as MHD winds or photoevaporation. These mechanisms must begin removing mass from the disks at a very early stage ($<0.5$\,Myr), as even for a region such as Ophiuchus, the masses of our viscous evolution models are larger than some of those observed. This hypothesis is also supported by \citet{AGEPRO_VII_population} and \citet{AGEPRO_VIII_ext_photoevap}, where they explore the role of MHD and (internal/external) photoevaporation in removing gas mass from the disk. 

Alternatively, external perturbers in the form of companions or fly-bys could also contribute to dispersing disk masses through tidal interactions \citep[see][for reviews]{offner2023, zurlo2023}, especially at early ages when SFRs have a higher number density of stars. Several systems from the literature have been reported as having gas morphologies consistent with simulations of interactions over the last few hundred of years \citep[e.g.,][]{kurtovic2018, Rodriguez2018, menard2020, zapata2020}, and systems that seem isolated today could have suffered from a fly-by tens of thousands of years ago or earlier \citep[e.g.,][]{winter2018}. Thus, the large range of $M_{\rm gas}$ observed in AGE-PRO might originate from a combination of internal and external mechanisms. 

Among the external mechanisms that could modify the mass of a disk, recent population synthesis simulations have shown that late infall of material could contribute to increasing the mass reservoir of a disk by more than 50\% \citep{kuffmeier2023} within a disk lifetime. These infalls have already been observed toward young objects \citep[e.g.,][]{yen2017, grant2021, cacciapuoti2024}. These positive mass contributions put a higher tension between the AGE-PRO observations and explaining mass loss through only viscous evolution models.

Despite the low values of $M_{\rm gas}$ that are not reproduced by our models, the comparison between $M_{\rm pebble}$ (or $F_{\rm mm}$) shows a better agreement, suggesting that early dispersal mechanisms at $<1$,Myr are affecting the gas more than the dust component. Tentatively, dust growth and evolution through radial drift could be protecting the dust reservoir at more compact radii in the midplane of disks compared to gas, which extends much farther from the star. Since $M_{\rm gas}$ is higher in our simulations, it implies that there are high dust-to-gas ratios in the observations that we do not reproduce in our models. The impact of this discrepancy on the trends derived from dust observables remains to be investigated.

\subsection{Trends in AGE-PRO that Are reproduced by dust evolution models}\label{sec:discussion:dust}

We find a better agreement when we compare the properties of dust continuum emission of the AGE-PRO observations to the dust evolution simulations with either strong or weak traps. The combination of disk size in continuum emission, as well as the continuum brightness, allows us to exclude disks with no pressure traps as the dominant population in the observations of AGE-PRO, in line with the previous results from \citet{zormpas2022} and \citet{delussu2024}. 

At the very early disk stages, traced by the sources in Ophiuchus, the observed values of $F_{mm}$ or $R_{90\%,mm}$ are consistent with any of the trapping scenarios, suggesting there has not been enough time for dust radial drift to deplete the disks of solid material. 
Thus, the solid content at early stages ($<$1\,Myr) might be a good tracer of the initial dust content of disks if the young disks have had a constant low viscosity in their midplanes. 
By the age of Lupus and Upper Sco, the disks are more consistent with the simulations with either weak or strong dust trapping, as those with no traps have already lost a significant amount of its solid reservoir. It is relevant to mention that our disks have static traps over time and location, which contributes to trapping dust more efficiently. Variability in the dust trapping strength or location would result in more leaky traps, which would make the results more similar to those of simulations without traps \citep{Pinilla2021_b}.

When comparing observations to radiative transfer models, it is relevant to consider that the contribution of small grains ($<100\,\mu$m) with the opacity from \citet{Ricci2010} is higher than that from other opacities, such as the DSHARP opacities (\citet{Birnstiel2018}, see, for example, \citet{Stadler2022}), thus the radiative transfer image of simulations with very little pebble mass ($<0.1\,M_\oplus$) can have detectable millimeter brightness, considering the AGE-PRO sensitivity. Even though the millimeter flux of simulations with no traps at 10\,Myr of age is comparable to fluxes of some Upper Sco disks, the $R_{90\%}$ of the simulations does not coincide with the observed radii. Further studies considering multimillimeter wavelength observations will be able to set stronger constraints on the grain size distribution \citep[e.g.,][]{sierra2024}, thus allowing for better direct comparisons between radiative transfer models and the AGE-PRO observations. 

Previous studies have shown that there is a decreasing trend in dust continuum millimeter brightness as a function of age \citep{hendler2020, manara2023}, which is also observed within the AGE-PRO sample (see Fig.~\ref{fig:mstar_1.3mmfluxes}). This decreasing trend with age is also reproduced in our simulations in $M_{\text{pebble}}$ or $F_{1.3mm}$, but it can be significantly slowed down for disks with strong traps, thus increasing the range of observed millimeter fluxes for a given SFR. The disks with high dust content and older than 3\,Myr favor the scenario of traps, when comparing their dust masses to those of the dust evolution simulations. 

When considering the evolution of size in disks, we find that the connection between $R_{\rm 90\%,pebble}$ and $R_{\rm 90\%,flux}$ has a high dependency on the choice of opacities, as also discussed by \citet{Stadler2022} and \citet{delussu2024}. Observationally, the sizes of disks are typically measured from millimeter continuum emission, and they are used as a tracer for dust mass distribution. However, these values can have large discrepancies if most of the dust mass is contained in small grains ($<100\,\mu$m). For example, the $R_{\rm 90\%,pebble}$ (which only considers grains $>100\,\mu$m) in disks without traps shows a monotonically decreasing trend as a function of age, which is a natural result of dust radial drift. However, when considering the $R_{90\%}$ of the disks in millimeter brightness, we see these disks becoming larger as a function of time, as the contributions from small grains in the outer part of the disks become the dominant millimeter flux source.

The distinction of pebbles used in this work is relevant to compare with previous works analyzing the evolution of disk sizes. For example, \citet{rosotti2019} found that the radius containing the total dust mass of the disk grows as a function of time in the absence of traps, as a result of micron-sized grains being coupled to the expanding gas in viscous evolution. This is in contrast to the behavior of our $R_{\rm 90\%, pebble}$ in the absence of traps, which does not consider the contribution of micron-sized grains to the pebble-mass content. Depending on the millimeter opacity of these small grains ($<100\,\mu$m-sized), the radiative transfer models of disks with no traps can show shrinking radii as a function of time \citep[as in ][]{rosotti2019}, or increasing radii as a function of time (this work; see Fig.~\ref{fig:mstar_rmod90_b6}); thus being dust opacity is one of the most relevant sources of uncertainty when comparing models to observations \citep[see also][]{delussu2024}. 

In the presence of traps, the connection between $R_{\rm 90\%, pebble}$ and $R_{\rm 90\%, mm}$ becomes more direct. The value of $R_{\rm 90\%, pebble}$ is completely determined by the location of the outermost dust trap, which is also true for $R_{90\%, mm}$ as long as the millimeter flux contribution of the material in the trap is larger than that of the micron-sized grains in the outer disk. Thus, the size of disks in millimeter emission might not be a good tracer of disk evolution, consistent with the findings of \citet{mulders2017}. In \citet{AGEPRO_V_gasmasses}, a positive correlation between $M_{\rm gas}$ and $R_{90\%,mm}$ is detected. If $R_{90\%,mm}$ is indeed tracing the location of the outermost trap, this relation could be evidence of more-massive disks having a higher potential for generating dust traps at larger radii. Additional observations at higher angular resolution across the whole AGE-PRO sample are needed to test this hypothesis. 

Finally, we observe a large spread in the spectral indices both in Lupus and in Upper Sco, as presented in Fig.~\ref{fig:mstar_specidx_gallery}. From our simulations, an increase in the spectral index is expected as a function of age. In the absence of traps, this increase happens very early in a disk age (about 2\,Myr), and it is only in the presence of dust traps that the spectral index distribution can have a distribution ranging between 2-4 for later ages, as observed by AGE-PRO. Even though the role of self-scattering, optical depth, and free-free emission needs to be further investigated with additional multiwavelength observations, our measurements of spectral indices offer an additional support for the presence of either weak or strong traps in the AGE-PRO sample.

%%%%%%%%%%%%%%%%%%%%%%%%%%%%%%%%%%%%%%%%%%%%%%
\section{Conclusions}\label{section_Summary}
%%%%%%%%%%%%%%%%%%%%%%%%%%%%%%%%%%%%%%%%%%%%%%

In this work, we compared dust evolution models to the AGE-PRO observations. Our models considered similar stellar masses and luminosities as the host stars of the AGE-PRO sample and included three different initial gas disk masses and four initial disk characteristic radii. Additionally, we assumed models that either do not have any dust traps or models that have weak or strong dust traps. The results of the dust evolution are taken to perform radiative transfer calculations, considering five different evolution times $t=[0.5, 1,2,5,10]$\,Myr. These outputs are compared to observations of disks in Ophiuchus (compared to output at 0.5\,Myr), Lupus (1 and 2\,Myr), and Upper Sco (2 and 5\,Myr). The main conclusions of our work are as follows: 

\begin{itemize}

    \item The pebble masses and millimeter fluxes obtained at 0.5\,Myr of evolution are representative of the initial dust reservoir for planet formation in disks, if the disks have had a low viscosity from the beginning. For this reason, when comparing our models to observations of the young disks in Ophiuchus, the data does not strongly favor models with or without substructures, as little dust loss has occurred due to drift at these early stages. 

    \item When comparing our models at more evolved stages with the Lupus and Upper Sco data, the observations (in particular, the observed millimeter fluxes and spectral indices) favor the presence of dust traps in disks, as the simulations with no traps are unable to follow the observational trends. The preference between weak and strong traps depends on the time of evolution when we compare the data, and also on each specific disk. Our models suggest that the dust content of the disks in Upper Sco have mostly survived until their current age due to the presence of traps. When traps are present in disks, moderate drift can still happen, decreasing the millimeter flux over time, but allowing $R_{\rm dust}$ to remain stable over time, as found in the AGE-PRO observations. 

    \item The dust disk size is determined by the location of the outermost dust trap, even at early evolution times ($<1$\,Myr). The AGE-PRO observations show a positive trend between the $R_{\rm 90\%,mm}$ and $M_{\rm gas}$. In the context of the models of this work, this would imply that more-massive disks have a higher potential to form traps further out. In the absence of traps, the expected shrink of $R_{\rm 90\%,mm}$ over time is not obtained in our synthetic observations: when drift removes millimeter-sized particles in the outer parts of the disks, the micron-sized particles start to dominate the millimeter emission, which could contribute to hiding the shrinking of disk size in high sensitivity observations, as it is the case for AGE-PRO. 

    \item The spatially integrated spectral index obtained for Lupus and Upper Sco sources in AGE-PRO shows a large spread, ranging from disks that are consistent with being completely optically thick, to others where the emission is more consistent with being optically thin at the wavelength of the AGE-PRO observations.  In simulations with stronger traps, the spectral index can remain closer to $\alpha_{\rm{mm}} \approx 2$ until later ages, compared to weaker traps. Thus, the spectral index comparison also favors the presence of traps. 

\end{itemize}

\begin{acknowledgments}

N.T.K. acknowledges the support of Ewine van Dishoeck while working on this project. The authors also thank the feedback provided by the referee, which contributed to improve this manuscript. 
N.T.K., M.G., and P.P. acknowledge the support provided by the Alexander von Humboldt Foundation in the framework of the Sofja Kovalevskaja Award endowed by the Federal Ministry of Education and Research. 
P.P. and A.S. acknowledge the support from the UK Research and Innovation (UKRI) under the UK government's Horizon Europe funding guarantee from ERC (under grant agreement No. 101076489).
K.Z. and L.T. acknowledge the support of the NSF AAG grant No. \#2205617. 
G.R. and R.A. acknowledge funding from the Fondazione Cariplo, grant No. 2022-1217, and the European Research Council (ERC) under the European Union's Horizon Europe Research \& Innovation Program under grant agreement No. 101039651 (DiscEvol). Views and opinions expressed are, however, those of the author(s) only, and do not necessarily reflect those of the European Union or the European Research Council Executive Agency. Neither the European Union nor the granting authority can be held responsible for them.
I.P. and D.D. acknowledge support from Collaborative NSF Astronomy \& Astrophysics Research grant (ID: 2205870).
B.T. acknowledges support from the Programme National “Physique et Chimie du Milieu Interstellaire” (PCMI) of CNRS/INSU with INC/INP and co-funded by CNES.
L.A.C, C.G.R., and J.M., acknowledge support from the Millennium Nucleus on Young Exoplanets and their Moons (YEMS), ANID - Center Code NCN2021\_080 and 
L.A.C. also acknowledges support from the FONDECYT grant \#1241056.
L.P. acknowledges support from ANID BASAL project FB210003 and ANID FONDECYT Regular \#1221442
C.A.G. and L.P. acknowledge support from FONDECYT de Postdoctorado 2021 \#3210520.
A.S. acknowledges support from FONDECYT de Postdoctorado 2022 $\#$3220495.
J.M acknowledges support from FONDECYT de Postdoctorado 2024 \#3240612. 

This paper makes use of the following ALMA data: ADS/JAO.ALMA\#2021.1.00128.L. ALMA is a partnership of ESO (representing its member states), NSF (USA) and NINS (Japan), together with NRC (Canada), MOST and ASIAA (Taiwan), and KASI (Republic of Korea), in cooperation with the Republic of Chile. The Joint ALMA Observatory is operated by ESO, AUI/NRAO and NAOJ. The National Radio Astronomy Observatory is a facility of the National Science Foundation operated under cooperative agreement by Associated Universities, Inc.

\end{acknowledgments}

\bibliographystyle{aasjournal}
\bibliography{TheBibliography}

\begin{thebibliography}{}
\expandafter\ifx\csname natexlab\endcsname\relax\def\natexlab#1{#1}\fi
\providecommand{\url}[1]{\href{#1}{#1}}

\bibitem[{{Agurto-Gangas} {et~al.}(2025){Agurto-Gangas}, {P{\'e}rez}, {Sierra}, {Miley}, {Zhang}, {Pascucci}, {Pinilla}, {Deng}, {Carpenter}, {Trapman}, {Vioque}, {Rosotti}, {Kurtovic}, {Cieza}, {Anania}, {Tabone}, {Schwarz}, {Hogerheijde}, {TorresVillanueva}, {Ruiz-Rodriguez}, \& {Gonz{\'a}lez-Ruilova}}]{AGEPRO_IV_UpperSco}
{Agurto-Gangas}, C., {P{\'e}rez}, L.~M., {Sierra}, A., {et~al.} 2025, \apj, 989, 4

\bibitem[{{Alexander} {et~al.}(2014){Alexander}, {Pascucci}, {Andrews}, {Armitage}, \& {Cieza}}]{Alexander2014}
{Alexander}, R., {Pascucci}, I., {Andrews}, S., {Armitage}, P., \& {Cieza}, L. 2014, in Protostars and Planets VI, ed. H.~{Beuther}, R.~S. {Klessen}, C.~P. {Dullemond}, \& T.~{Henning}, 475--496

\bibitem[{{Anania} {et~al.}(2025){Anania}, {Rosotti}, {G{\'a}rate}, {Pinilla}, {Vioque}, {Trapman}, {Carpenter}, {Zhang}, {Pascucci}, {Cieza}, {Sierra}, {Kurtovic}, {Miley}, {P{\'e}rez}, {Tabone}, {Hogerheijde}, {Deng}, {Agurto-Gangas}, {Ruiz-Rodriguez}, {Gonz{\'a}lez-Ruilova}, \& {TorresVillanueva}}]{AGEPRO_VIII_ext_photoevap}
{Anania}, R., {Rosotti}, G.~P., {G{\'a}rate}, M., {et~al.} 2025, \apj, 989, 8

\bibitem[{{Andrews}(2020)}]{Andrews2020_review}
{Andrews}, S.~M. 2020, \araa, 58, 483

\bibitem[{{Andrews} {et~al.}(2013){Andrews}, {Rosenfeld}, {Kraus}, \& {Wilner}}]{andrews2013}
{Andrews}, S.~M., {Rosenfeld}, K.~A., {Kraus}, A.~L., \& {Wilner}, D.~J. 2013, \apj, 771, 129

\bibitem[{{Ansdell} {et~al.}(2016){Ansdell}, {Williams}, {van der Marel}, {Carpenter}, {Guidi}, {Hogerheijde}, {Mathews}, {Manara}, {Miotello}, {Natta}, {Oliveira}, {Tazzari}, {Testi}, {van Dishoeck}, \& {van Terwisga}}]{Ansdell2016}
{Ansdell}, M., {Williams}, J.~P., {van der Marel}, N., {et~al.} 2016, \apj, 828, 46

\bibitem[{{Bae} {et~al.}(2023){Bae}, {Isella}, {Zhu}, {Martin}, {Okuzumi}, \& {Suriano}}]{Bae2023}
{Bae}, J., {Isella}, A., {Zhu}, Z., {et~al.} 2023, in Astronomical Society of the Pacific Conference Series, Vol. 534, Protostars and Planets VII, ed. S.~{Inutsuka}, Y.~{Aikawa}, T.~{Muto}, K.~{Tomida}, \& M.~{Tamura}, 423

\bibitem[{{Barenfeld} {et~al.}(2016){Barenfeld}, {Carpenter}, {Ricci}, \& {Isella}}]{barenfeld2016}
{Barenfeld}, S.~A., {Carpenter}, J.~M., {Ricci}, L., \& {Isella}, A. 2016, \apj, 827, 142

\bibitem[{{Bate}(2018)}]{Bate2018}
{Bate}, M.~R. 2018, \mnras, 475, 5618

\bibitem[{{Birnstiel}(2024)}]{birnstiel2023}
{Birnstiel}, T. 2024, \araa, 62, 157

\bibitem[{{Birnstiel} {et~al.}(2010){Birnstiel}, {Dullemond}, \& {Brauer}}]{Birnstiel2010}
{Birnstiel}, T., {Dullemond}, C.~P., \& {Brauer}, F. 2010, \aap, 513, A79

\bibitem[{{Birnstiel} {et~al.}(2016){Birnstiel}, {Fang}, \& {Johansen}}]{Birnstiel2016_Review}
{Birnstiel}, T., {Fang}, M., \& {Johansen}, A. 2016, \ssr, 205, 41

\bibitem[{{Birnstiel} {et~al.}(2018){Birnstiel}, {Dullemond}, {Zhu}, {Andrews}, {Bai}, {Wilner}, {Carpenter}, {Huang}, {Isella}, {Benisty}, {P{\'e}rez}, \& {Zhang}}]{Birnstiel2018}
{Birnstiel}, T., {Dullemond}, C.~P., {Zhu}, Z., {et~al.} 2018, \apjl, 869, L45

\bibitem[{{Cacciapuoti} {et~al.}(2024){Cacciapuoti}, {Macias}, {Gupta}, {Testi}, {Miotello}, {Espaillat}, {K{\"u}ffmeier}, {van Terwisga}, {Tobin}, {Grant}, {Manara}, {Segura-Cox}, {Wendeborn}, {Klessen}, {Maury}, {Lebreuilly}, {Hennebelle}, \& {Molinari}}]{cacciapuoti2024}
{Cacciapuoti}, L., {Macias}, E., {Gupta}, A., {et~al.} 2024, \aap, 682, A61

\bibitem[{{Clark}(1980)}]{clark1980}
{Clark}, B.~G. 1980, \aap, 89, 377

\bibitem[{{Clarke} {et~al.}(2001){Clarke}, {Gendrin}, \& {Sotomayor}}]{Clarke2001}
{Clarke}, C.~J., {Gendrin}, A., \& {Sotomayor}, M. 2001, \mnras, 328, 485

\bibitem[{{Das} {et~al.}(2024){Das}, {Kurtovic}, \& {Flock}}]{das2024}
{Das}, S., {Kurtovic}, N.~T., \& {Flock}, M. 2024, \aap, 689, A104

\bibitem[{{Delage} {et~al.}(2023){Delage}, {G{\'a}rate}, {Okuzumi}, {Yang}, {Pinilla}, {Flock}, {Stammler}, \& {Birnstiel}}]{delage2023}
{Delage}, T.~N., {G{\'a}rate}, M., {Okuzumi}, S., {et~al.} 2023, \aap, 674, A190

\bibitem[{{Delage} {et~al.}(2022){Delage}, {Okuzumi}, {Flock}, {Pinilla}, \& {Dzyurkevich}}]{Delage2022}
{Delage}, T.~N., {Okuzumi}, S., {Flock}, M., {Pinilla}, P., \& {Dzyurkevich}, N. 2022, \aap, 658, A97

\bibitem[{{Delussu} {et~al.}(2024){Delussu}, {Birnstiel}, {Miotello}, {Pinilla}, {Rosotti}, \& {Andrews}}]{delussu2024}
{Delussu}, L., {Birnstiel}, T., {Miotello}, A., {et~al.} 2024, \aap, 688, A81

\bibitem[{{Deng} {et~al.}(2025){Deng}, {Vioque}, {Pascucci}, {P{\'e}rez}, {Zhang}, {Kurtovic}, {Trapman}, {TorresVillanueva}, {Agurto-Gangas}, {Carpenter}, {Pinilla}, {Gorti}, {Tabone}, {Sierra}, {Rosotti}, {Cieza}, {Anania}, {Gonz{\'a}lez-Ruilova}, {Hogerheijde}, {Miley}, {Ruiz-Rodriguez}, {Ruaud}, \& {Schwarz}}]{AGEPRO_III_Lupus}
{Deng}, D., {Vioque}, M., {Pascucci}, I., {et~al.} 2025, \apj, 989, 3

\bibitem[{{Doi} \& {Kataoka}(2023)}]{doi2023}
{Doi}, K., \& {Kataoka}, A. 2023, \apj, 957, 11

\bibitem[{{Dominik} {et~al.}(2021){Dominik}, {Min}, \& {Tazaki}}]{Dominik2021}
{Dominik}, C., {Min}, M., \& {Tazaki}, R. 2021, {OpTool: Command-line driven tool for creating complex dust opacities}, , , ascl:2104.010

\bibitem[{{Dotter} {et~al.}(2008){Dotter}, {Chaboyer}, {Jevremovi{\'c}}, {Kostov}, {Baron}, \& {Ferguson}}]{dotter2008}
{Dotter}, A., {Chaboyer}, B., {Jevremovi{\'c}}, D., {et~al.} 2008, \apjs, 178, 89

\bibitem[{{Draine}(2003)}]{Draine2003}
{Draine}, B.~T. 2003, \araa, 41, 241

\bibitem[{{Dr{\c a}{\.z}kowska} {et~al.}(2023){Dr{\c a}{\.z}kowska}, {Bitsch}, {Lambrechts}, {Mulders}, {Harsono}, {Vazan}, {Liu}, {Ormel}, {Kretke}, \& {Morbidelli}}]{Drazkowska2023}
{Dr{\c a}{\.z}kowska}, J., {Bitsch}, B., {Lambrechts}, M., {et~al.} 2023, in Astronomical Society of the Pacific Conference Series, Vol. 534, Protostars and Planets VII, ed. S.~{Inutsuka}, Y.~{Aikawa}, T.~{Muto}, K.~{Tomida}, \& M.~{Tamura}, 717

\bibitem[{{Dubrulle} {et~al.}(1995){Dubrulle}, {Morfill}, \& {Sterzik}}]{Dubrelle1995}
{Dubrulle}, B., {Morfill}, G., \& {Sterzik}, M. 1995, \icarus, 114, 237

\bibitem[{{Dullemond} {et~al.}(2012){Dullemond}, {Juhasz}, {Pohl}, {Sereshti}, {Shetty}, {Peters}, {Commercon}, \& {Flock}}]{RADMC3D2012}
{Dullemond}, C.~P., {Juhasz}, A., {Pohl}, A., {et~al.} 2012, {RADMC-3D: A multi-purpose radiative transfer tool}, , , ascl:1202.015

\bibitem[{{Fedele} {et~al.}(2010){Fedele}, {van den Ancker}, {Henning}, {Jayawardhana}, \& {Oliveira}}]{Fedele2010}
{Fedele}, D., {van den Ancker}, M.~E., {Henning}, T., {Jayawardhana}, R., \& {Oliveira}, J.~M. 2010, \aap, 510, A72

\bibitem[{{Fromang} \& {Nelson}(2009)}]{Fromang2009}
{Fromang}, S., \& {Nelson}, R.~P. 2009, \aap, 496, 597

\bibitem[{{Gaia Collaboration} {et~al.}(2021){Gaia Collaboration}, {Brown}, {Vallenari}, {Prusti}, {de Bruijne}, {Babusiaux}, {Biermann}, {Creevey}, {Evans}, {Eyer}, {Hutton}, {Jansen}, {Jordi}, {Klioner}, {Lammers}, {Lindegren}, {Luri}, {Mignard}, {Panem}, {Pourbaix}, {Randich}, {Sartoretti}, {Soubiran}, {Walton}, {Arenou}, {Bailer-Jones}, {Bastian}, {Cropper}, {Drimmel}, {Katz}, {Lattanzi}, {van Leeuwen}, {Bakker}, {Cacciari}, {Casta{\~n}eda}, {De Angeli}, {Ducourant}, {Fabricius}, {Fouesneau}, {Fr{\'e}mat}, {Guerra}, {Guerrier}, {Guiraud}, {Jean-Antoine Piccolo}, {Masana}, {Messineo}, {Mowlavi}, {Nicolas}, {Nienartowicz}, {Pailler}, {Panuzzo}, {Riclet}, {Roux}, {Seabroke}, {Sordo}, {Tanga}, {Th{\'e}venin}, {Gracia-Abril}, {Portell}, {Teyssier}, {Altmann}, {Andrae}, {Bellas-Velidis}, {Benson}, {Berthier}, {Blomme}, {Brugaletta}, {Burgess}, {Busso}, {Carry}, {Cellino}, {Cheek}, {Clementini}, {Damerdji}, {Davidson}, {Delchambre}, {Dell'Oro}, {Fern{\'a}ndez-Hern{\'a}ndez}, {Galluccio}, {Garc{\'\i}a-Lario},
  {Garcia-Reinaldos}, {Gonz{\'a}lez-N{\'u}{\~n}ez}, {Gosset}, {Haigron}, {Halbwachs}, {Hambly}, {Harrison}, {Hatzidimitriou}, {Heiter}, {Hern{\'a}ndez}, {Hestroffer}, {Hodgkin}, {Holl}, {Jan{\ss}en}, {Jevardat de Fombelle}, {Jordan}, {Krone-Martins}, {Lanzafame}, {L{\"o}ffler}, {Lorca}, {Manteiga}, {Marchal}, {Marrese}, {Moitinho}, {Mora}, {Muinonen}, {Osborne}, {Pancino}, {Pauwels}, {Petit}, {Recio-Blanco}, {Richards}, {Riello}, {Rimoldini}, {Robin}, {Roegiers}, {Rybizki}, {Sarro}, {Siopis}, {Smith}, {Sozzetti}, {Ulla}, {Utrilla}, {van Leeuwen}, {van Reeven}, {Abbas}, {Abreu Aramburu}, {Accart}, {Aerts}, {Aguado}, {Ajaj}, {Altavilla}, {{\'A}lvarez}, {{\'A}lvarez Cid-Fuentes}, {Alves}, {Anderson}, {Anglada Varela}, {Antoja}, {Audard}, {Baines}, {Baker}, {Balaguer-N{\'u}{\~n}ez}, {Balbinot}, {Balog}, {Barache}, {Barbato}, {Barros}, {Barstow}, {Bartolom{\'e}}, {Bassilana}, {Bauchet}, {Baudesson-Stella}, {Becciani}, {Bellazzini}, {Bernet}, {Bertone}, {Bianchi}, {Blanco-Cuaresma}, {Boch}, {Bombrun}, {Bossini},
  {Bouquillon}, {Bragaglia}, {Bramante}, {Breedt}, {Bressan}, {Brouillet}, {Bucciarelli}, {Burlacu}, {Busonero}, {Butkevich}, {Buzzi}, {Caffau}, {Cancelliere}, {C{\'a}novas}, {Cantat-Gaudin}, {Carballo}, {Carlucci}, {Carnerero}, {Carrasco}, {Casamiquela}, {Castellani}, {Castro-Ginard}, {Castro Sampol}, {Chaoul}, {Charlot}, {Chemin}, {Chiavassa}, {Cioni}, {Comoretto}, {Cooper}, {Cornez}, {Cowell}, {Crifo}, {Crosta}, {Crowley}, {Dafonte}, {Dapergolas}, {David}, {David}, {de Laverny}, {De Luise}, {De March}, {De Ridder}, {de Souza}, {de Teodoro}, {de Torres}, {del Peloso}, {del Pozo}, {Delbo}, {Delgado}, {Delgado}, {Delisle}, {Di Matteo}, {Diakite}, {Diener}, {Distefano}, {Dolding}, {Eappachen}, {Edvardsson}, {Enke}, {Esquej}, {Fabre}, {Fabrizio}, {Faigler}, {Fedorets}, {Fernique}, {Fienga}, {Figueras}, {Fouron}, {Fragkoudi}, {Fraile}, {Franke}, {Gai}, {Garabato}, {Garcia-Gutierrez}, {Garc{\'\i}a-Torres}, {Garofalo}, {Gavras}, {Gerlach}, {Geyer}, {Giacobbe}, {Gilmore}, {Girona}, {Giuffrida}, {Gomel}, {Gomez},
  {Gonzalez-Santamaria}, {Gonz{\'a}lez-Vidal}, {Granvik}, {Guti{\'e}rrez-S{\'a}nchez}, {Guy}, {Hauser}, {Haywood}, {Helmi}, {Hidalgo}, {Hilger}, {H{\l}adczuk}, {Hobbs}, {Holland}, {Huckle}, {Jasniewicz}, {Jonker}, {Juaristi Campillo}, {Julbe}, {Karbevska}, {Kervella}, {Khanna}, {Kochoska}, {Kontizas}, {Kordopatis}, {Korn}, {Kostrzewa-Rutkowska}, {Kruszy{\'n}ska}, {Lambert}, {Lanza}, {Lasne}, {Le Campion}, {Le Fustec}, {Lebreton}, {Lebzelter}, {Leccia}, {Leclerc}, {Lecoeur-Taibi}, {Liao}, {Licata}, {Lindstr{\o}m}, {Lister}, {Livanou}, {Lobel}, {Madrero Pardo}, {Managau}, {Mann}, {Marchant}, {Marconi}, {Marcos Santos}, {Marinoni}, {Marocco}, {Marshall}, {Martin Polo}, {Mart{\'\i}n-Fleitas}, {Masip}, {Massari}, {Mastrobuono-Battisti}, {Mazeh}, {McMillan}, {Messina}, {Michalik}, {Millar}, {Mints}, {Molina}, {Molinaro}, {Moln{\'a}r}, {Montegriffo}, {Mor}, {Morbidelli}, {Morel}, {Morris}, {Mulone}, {Munoz}, {Muraveva}, {Murphy}, {Musella}, {Noval}, {Ord{\'e}novic}, {Orr{\`u}}, {Osinde}, {Pagani}, {Pagano},
  {Palaversa}, {Palicio}, {Panahi}, {Pawlak}, {Pe{\~n}alosa Esteller}, {Penttil{\"a}}, {Piersimoni}, {Pineau}, {Plachy}, {Plum}, {Poggio}, {Poretti}, {Poujoulet}, {Pr{\v{s}}a}, {Pulone}, {Racero}, {Ragaini}, {Rainer}, {Raiteri}, {Rambaux}, {Ramos}, {Ramos-Lerate}, {Re Fiorentin}, {Regibo}, {Reyl{\'e}}, {Ripepi}, {Riva}, {Rixon}, {Robichon}, {Robin}, {Roelens}, {Rohrbasser}, {Romero-G{\'o}mez}, {Rowell}, {Royer}, {Rybicki}, {Sadowski}, {Sagrist{\`a} Sell{\'e}s}, {Sahlmann}, {Salgado}, {Salguero}, {Samaras}, {Sanchez Gimenez}, {Sanna}, {Santove{\~n}a}, {Sarasso}, {Schultheis}, {Sciacca}, {Segol}, {Segovia}, {S{\'e}gransan}, {Semeux}, {Shahaf}, {Siddiqui}, {Siebert}, {Siltala}, {Slezak}, {Smart}, {Solano}, {Solitro}, {Souami}, {Souchay}, {Spagna}, {Spoto}, {Steele}, {Steidelm{\"u}ller}, {Stephenson}, {S{\"u}veges}, {Szabados}, {Szegedi-Elek}, {Taris}, {Tauran}, {Taylor}, {Teixeira}, {Thuillot}, {Tonello}, {Torra}, {Torra}, {Turon}, {Unger}, {Vaillant}, {van Dillen}, {Vanel}, {Vecchiato}, {Viala}, {Vicente},
  {Voutsinas}, {Weiler}, {Wevers}, {Wyrzykowski}, {Yoldas}, {Yvard}, {Zhao}, {Zorec}, {Zucker}, {Zurbach}, \& {Zwitter}}]{gaia2021edr3}
{Gaia Collaboration}, {Brown}, A.~G.~A., {Vallenari}, A., {et~al.} 2021, \aap, 649, A1

\bibitem[{{G{\'a}rate} {et~al.}(2020){G{\'a}rate}, {Birnstiel}, {Dr{\c{a}}{\.z}kowska}, \& {Stammler}}]{Garate2020}
{G{\'a}rate}, M., {Birnstiel}, T., {Dr{\c{a}}{\.z}kowska}, J., \& {Stammler}, S.~M. 2020, \aap, 635, A149

\bibitem[{{G{\'a}rate} {et~al.}(2021){G{\'a}rate}, {Delage}, {Stadler}, {Pinilla}, {Birnstiel}, {Stammler}, {Picogna}, {Ercolano}, {Franz}, \& {Lenz}}]{Garate2021}
{G{\'a}rate}, M., {Delage}, T.~N., {Stadler}, J., {et~al.} 2021, \aap, 655, A18

\bibitem[{{G{\'a}rate} {et~al.}(2023){G{\'a}rate}, {Birnstiel}, {Pinilla}, {Andrews}, {Franz}, {Stammler}, {Picogna}, {Ercolano}, {Miotello}, \& {Kurtovic}}]{garate2023}
{G{\'a}rate}, M., {Birnstiel}, T., {Pinilla}, P., {et~al.} 2023, \aap, 679, A15

\bibitem[{{Garufi} {et~al.}(2025){Garufi}, {Carrasco-Gonz{\'a}lez}, {Mac{\'\i}as}, {Testi}, {Curone}, {Ricci}, {Facchini}, {Long}, {Manara}, {Pascucci}, {Rosotti}, {Zagaria}, {Clarke}, {Herczeg}, {Isella}, {Rota}, {Mauc{\'o}}, {van der Marel}, \& {Tazzari}}]{garufi2025}
{Garufi}, A., {Carrasco-Gonz{\'a}lez}, C., {Mac{\'\i}as}, E., {et~al.} 2025, \aap, 694, A290

\bibitem[{{Grant} {et~al.}(2021){Grant}, {Espaillat}, {Wendeborn}, {Tobin}, {Mac{\'\i}as}, {Rilinger}, {Ribas}, {Megeath}, {Fischer}, {Calvet}, \& {Hee Kim}}]{grant2021}
{Grant}, S.~L., {Espaillat}, C.~C., {Wendeborn}, J., {et~al.} 2021, \apj, 913, 123

\bibitem[{{Gundlach} \& {Blum}(2015)}]{Gundlach2015}
{Gundlach}, B., \& {Blum}, J. 2015, \apj, 798, 34

\bibitem[{{Gundlach} {et~al.}(2011){Gundlach}, {Kilias}, {Beitz}, \& {Blum}}]{Gundlach2011}
{Gundlach}, B., {Kilias}, S., {Beitz}, E., \& {Blum}, J. 2011, \icarus, 214, 717

\bibitem[{{Gundlach} {et~al.}(2018){Gundlach}, {Schmidt}, {Kreuzig}, {Bischoff}, {Rezaei}, {Kothe}, {Blum}, {Grzesik}, \& {Stoll}}]{Gundlach2018}
{Gundlach}, B., {Schmidt}, K.~P., {Kreuzig}, C., {et~al.} 2018, \mnras, 479, 1273

\bibitem[{{Hendler} {et~al.}(2020){Hendler}, {Pascucci}, {Pinilla}, {Tazzari}, {Carpenter}, {Malhotra}, \& {Testi}}]{hendler2020}
{Hendler}, N., {Pascucci}, I., {Pinilla}, P., {et~al.} 2020, \apj, 895, 126

\bibitem[{{Hildebrand}(1983)}]{hildebrand1983}
{Hildebrand}, R.~H. 1983, \qjras, 24, 267

\bibitem[{{H{\"o}gbom}(1974)}]{hogbom1974}
{H{\"o}gbom}, J.~A. 1974, \aaps, 15, 417

\bibitem[{{Huang} {et~al.}(2018){Huang}, {Andrews}, {Cleeves}, {{\"O}berg}, {Wilner}, {Bai}, {Birnstiel}, {Carpenter}, {Hughes}, {Isella}, {P{\'e}rez}, {Ricci}, \& {Zhu}}]{Huang2018}
{Huang}, J., {Andrews}, S.~M., {Cleeves}, L.~I., {et~al.} 2018, \apj, 852, 122

\bibitem[{{Kuffmeier} {et~al.}(2023){Kuffmeier}, {Jensen}, \& {Haugb{\o}lle}}]{kuffmeier2023}
{Kuffmeier}, M., {Jensen}, S.~S., \& {Haugb{\o}lle}, T. 2023, European Physical Journal Plus, 138, 272

\bibitem[{{Kurtovic} {et~al.}(2018){Kurtovic}, {P{\'e}rez}, {Benisty}, {Zhu}, {Zhang}, {Huang}, {Andrews}, {Dullemond}, {Isella}, {Bai}, {Carpenter}, {Guzm{\'a}n}, {Ricci}, \& {Wilner}}]{kurtovic2018}
{Kurtovic}, N.~T., {P{\'e}rez}, L.~M., {Benisty}, M., {et~al.} 2018, \apjl, 869, L44

\bibitem[{{Kurtovic} {et~al.}(2021){Kurtovic}, {Pinilla}, {Long}, {Benisty}, {Manara}, {Natta}, {Pascucci}, {Ricci}, {Scholz}, \& {Testi}}]{kurtovic2021}
{Kurtovic}, N.~T., {Pinilla}, P., {Long}, F., {et~al.} 2021, \aap, 645, A139

\bibitem[{{Lesur} {et~al.}(2023){Lesur}, {Flock}, {Ercolano}, {Lin}, {Yang}, {Barranco}, {Benitez-Llambay}, {Goodman}, {Johansen}, {Klahr}, {Laibe}, {Lyra}, {Marcus}, {Nelson}, {Squire}, {Simon}, {Turner}, {Umurhan}, \& {Youdin}}]{lesur2023}
{Lesur}, G., {Flock}, M., {Ercolano}, B., {et~al.} 2023, in Astronomical Society of the Pacific Conference Series, Vol. 534, Protostars and Planets VII, ed. S.~{Inutsuka}, Y.~{Aikawa}, T.~{Muto}, K.~{Tomida}, \& M.~{Tamura}, 465

\bibitem[{{Liu}(2019)}]{liu2019}
{Liu}, H.~B. 2019, \apjl, 877, L22

\bibitem[{{Lodato} {et~al.}(2017){Lodato}, {Scardoni}, {Manara}, \& {Testi}}]{lodato2017}
{Lodato}, G., {Scardoni}, C.~E., {Manara}, C.~F., \& {Testi}, L. 2017, \mnras, 472, 4700

\bibitem[{{Long} {et~al.}(2018){Long}, {Pinilla}, {Herczeg}, {Harsono}, {Dipierro}, {Pascucci}, {Hendler}, {Tazzari}, {Ragusa}, {Salyk}, {Edwards}, {Lodato}, {van de Plas}, {Johnstone}, {Liu}, {Boehler}, {Cabrit}, {Manara}, {Menard}, {Mulders}, {Nisini}, {Fischer}, {Rigliaco}, {Banzatti}, {Avenhaus}, \& {Gully-Santiago}}]{long2018}
{Long}, F., {Pinilla}, P., {Herczeg}, G.~J., {et~al.} 2018, \apj, 869, 17

\bibitem[{{Long} {et~al.}(2020){Long}, {Pinilla}, {Herczeg}, {Andrews}, {Harsono}, {Johnstone}, {Ragusa}, {Pascucci}, {Wilner}, {Hendler}, {Jennings}, {Liu}, {Lodato}, {Menard}, {van de Plas}, \& {Dipierro}}]{long2020}
---. 2020, \apj, 898, 36

\bibitem[{{Long} {et~al.}(2022){Long}, {Andrews}, {Rosotti}, {Harsono}, {Pinilla}, {Wilner}, {{\"O}berg}, {Teague}, {Trapman}, \& {Tabone}}]{long2022}
{Long}, F., {Andrews}, S.~M., {Rosotti}, G., {et~al.} 2022, \apj, 931, 6

\bibitem[{{Lynden-Bell} \& {Pringle}(1974)}]{Lynden-Bell1974}
{Lynden-Bell}, D., \& {Pringle}, J.~E. 1974, \mnras, 168, 603

\bibitem[{{Manara} {et~al.}(2023){Manara}, {Ansdell}, {Rosotti}, {Hughes}, {Armitage}, {Lodato}, \& {Williams}}]{manara2023}
{Manara}, C.~F., {Ansdell}, M., {Rosotti}, G.~P., {et~al.} 2023, in Astronomical Society of the Pacific Conference Series, Vol. 534, Protostars and Planets VII, ed. S.~{Inutsuka}, Y.~{Aikawa}, T.~{Muto}, K.~{Tomida}, \& M.~{Tamura}, 539

\bibitem[{{Manara} {et~al.}(2016){Manara}, {Rosotti}, {Testi}, {Natta}, {Alcal{\'a}}, {Williams}, {Ansdell}, {Miotello}, {van der Marel}, {Tazzari}, {Carpenter}, {Guidi}, {Mathews}, {Oliveira}, {Prusti}, \& {van Dishoeck}}]{Manara2016_a}
{Manara}, C.~F., {Rosotti}, G., {Testi}, L., {et~al.} 2016, \aap, 591, L3

\bibitem[{{Mathis} {et~al.}(1977){Mathis}, {Rumpl}, \& {Nordsieck}}]{Mathis1977}
{Mathis}, J.~S., {Rumpl}, W., \& {Nordsieck}, K.~H. 1977, \apj, 217, 425

\bibitem[{{M{\'e}nard} {et~al.}(2020){M{\'e}nard}, {Cuello}, {Ginski}, {van der Plas}, {Villenave}, {Gonzalez}, {Pinte}, {Benisty}, {Boccaletti}, {Price}, {Boehler}, {Chripko}, {de Boer}, {Dominik}, {Garufi}, {Gratton}, {Hagelberg}, {Henning}, {Langlois}, {Maire}, {Pinilla}, {Ruane}, {Schmid}, {van Holstein}, {Vigan}, {Zurlo}, {Hubin}, {Pavlov}, {Rochat}, {Sauvage}, \& {Stadler}}]{menard2020}
{M{\'e}nard}, F., {Cuello}, N., {Ginski}, C., {et~al.} 2020, \aap, 639, L1

\bibitem[{{Mulders} {et~al.}(2017){Mulders}, {Pascucci}, {Manara}, {Testi}, {Herczeg}, {Henning}, {Mohanty}, \& {Lodato}}]{mulders2017}
{Mulders}, G.~D., {Pascucci}, I., {Manara}, C.~F., {et~al.} 2017, \apj, 847, 31

\bibitem[{{Musiolik} \& {Wurm}(2019)}]{Musiolik2019}
{Musiolik}, G., \& {Wurm}, G. 2019, \apj, 873, 58

\bibitem[{{Nakagawa} {et~al.}(1986){Nakagawa}, {Sekiya}, \& {Hayashi}}]{Nakagawa1986}
{Nakagawa}, Y., {Sekiya}, M., \& {Hayashi}, C. 1986, \icarus, 67, 375

\bibitem[{{Offner} {et~al.}(2023){Offner}, {Moe}, {Kratter}, {Sadavoy}, {Jensen}, \& {Tobin}}]{offner2023}
{Offner}, S.~S.~R., {Moe}, M., {Kratter}, K.~M., {et~al.} 2023, in Astronomical Society of the Pacific Conference Series, Vol. 534, Protostars and Planets VII, ed. S.~{Inutsuka}, Y.~{Aikawa}, T.~{Muto}, K.~{Tomida}, \& M.~{Tamura}, 275

\bibitem[{{Pascucci} {et~al.}(2023){Pascucci}, {Cabrit}, {Edwards}, {Gorti}, {Gressel}, \& {Suzuki}}]{pascucci2023}
{Pascucci}, I., {Cabrit}, S., {Edwards}, S., {et~al.} 2023, in Astronomical Society of the Pacific Conference Series, Vol. 534, Protostars and Planets VII, ed. S.~{Inutsuka}, Y.~{Aikawa}, T.~{Muto}, K.~{Tomida}, \& M.~{Tamura}, 567

\bibitem[{{Pascucci} {et~al.}(2016){Pascucci}, {Testi}, {Herczeg}, {Long}, {Manara}, {Hendler}, {Mulders}, {Krijt}, {Ciesla}, {Henning}, {Mohanty}, {Drabek-Maunder}, {Apai}, {Sz{\H{u}}cs}, {Sacco}, \& {Olofsson}}]{pascucci2016}
{Pascucci}, I., {Testi}, L., {Herczeg}, G.~J., {et~al.} 2016, \apj, 831, 125

\bibitem[{{Pinilla} {et~al.}(2012){Pinilla}, {Benisty}, \& {Birnstiel}}]{Pinilla2012}
{Pinilla}, P., {Benisty}, M., \& {Birnstiel}, T. 2012, \aap, 545, A81

\bibitem[{{Pinilla} {et~al.}(2013){Pinilla}, {Birnstiel}, {Benisty}, {Ricci}, {Natta}, {Dullemond}, {Dominik}, \& {Testi}}]{Pinilla2013}
{Pinilla}, P., {Birnstiel}, T., {Benisty}, M., {et~al.} 2013, \aap, 554, A95

\bibitem[{{Pinilla} {et~al.}(2021{\natexlab{a}}){Pinilla}, {Lenz}, \& {Stammler}}]{Pinilla2021_b}
{Pinilla}, P., {Lenz}, C.~T., \& {Stammler}, S.~M. 2021{\natexlab{a}}, \aap, 645, A70

\bibitem[{{Pinilla} {et~al.}(2020){Pinilla}, {Pascucci}, \& {Marino}}]{Pinilla2020}
{Pinilla}, P., {Pascucci}, I., \& {Marino}, S. 2020, \aap, 635, A105

\bibitem[{{Pinilla} {et~al.}(2021{\natexlab{b}}){Pinilla}, {Kurtovic}, {Benisty}, {Manara}, {Natta}, {Sanchis}, {Tazzari}, {Stammler}, {Ricci}, \& {Testi}}]{Pinilla2021}
{Pinilla}, P., {Kurtovic}, N.~T., {Benisty}, M., {et~al.} 2021{\natexlab{b}}, \aap, 649, A122

\bibitem[{{Pringle}(1981)}]{Pringle1981}
{Pringle}, J.~E. 1981, \araa, 19, 137

\bibitem[{{Ribas} {et~al.}(2015){Ribas}, {Bouy}, \& {Mer{\'\i}n}}]{Ribas2015}
{Ribas}, {\'A}., {Bouy}, H., \& {Mer{\'\i}n}, B. 2015, \aap, 576, A52

\bibitem[{{Ricci} {et~al.}(2010){Ricci}, {Testi}, {Natta}, {Neri}, {Cabrit}, \& {Herczeg}}]{Ricci2010}
{Ricci}, L., {Testi}, L., {Natta}, A., {et~al.} 2010, \aap, 512, A15

\bibitem[{{Rodriguez} {et~al.}(2018){Rodriguez}, {Loomis}, {Cabrit}, {Haworth}, {Facchini}, {Dougados}, {Booth}, {Jensen}, {Clarke}, {Stassun}, {Dent}, \& {Pety}}]{Rodriguez2018}
{Rodriguez}, J.~E., {Loomis}, R., {Cabrit}, S., {et~al.} 2018, \apj, 859, 150

\bibitem[{{Rosotti}(2023)}]{rosotti2023}
{Rosotti}, G.~P. 2023, \nar, 96, 101674

\bibitem[{{Rosotti} {et~al.}(2019){Rosotti}, {Tazzari}, {Booth}, {Testi}, {Lodato}, \& {Clarke}}]{rosotti2019}
{Rosotti}, G.~P., {Tazzari}, M., {Booth}, R.~A., {et~al.} 2019, \mnras, 486, 4829

\bibitem[{{Ruiz-Rodriguez} {et~al.}(2025){Ruiz-Rodriguez}, {Gonz{\'a}lez-Ruilova}, {Cieza}, {Zhang}, {Trapman}, {Sierra}, {Pinilla}, {Pascucci}, {P{\'e}rez}, {Deng}, {Agurto-Gangas}, {Carpenter}, {Tabone}, {Rosotti}, {Anania}, {Miley}, {Schwarz}, {Kuznetsova}, {Vioque}, \& {Kurtovic}}]{AGEPRO_II_Ophiuchus}
{Ruiz-Rodriguez}, D.~A., {Gonz{\'a}lez-Ruilova}, C., {Cieza}, L.~A., {et~al.} 2025, \apj, 989, 2

\bibitem[{{Sanchis} {et~al.}(2021){Sanchis}, {Testi}, {Natta}, {Facchini}, {Manara}, {Miotello}, {Ercolano}, {Henning}, {Preibisch}, {Carpenter}, {de Gregorio-Monsalvo}, {Jayawardhana}, {Lopez}, {Mu{\v{z}}i{\'c}}, {Pascucci}, {Santamar{\'\i}a-Miranda}, {van Terwisga}, \& {Williams}}]{sanchis2021}
{Sanchis}, E., {Testi}, L., {Natta}, A., {et~al.} 2021, \aap, 649, A19

\bibitem[{{Shakura} \& {Sunyaev}(1973)}]{Shakura1973}
{Shakura}, N.~I., \& {Sunyaev}, R.~A. 1973, \aap, 24, 337

\bibitem[{{Sierra} {et~al.}(2024){Sierra}, {P{\'e}rez}, {Sotomayor}, {Benisty}, {Chandler}, {Andrews}, {Carpenter}, {Henning}, {Testi}, {Ricci}, \& {Wilner}}]{sierra2024}
{Sierra}, A., {P{\'e}rez}, L.~M., {Sotomayor}, B., {et~al.} 2024, \apj, 974, 306

\bibitem[{{Somigliana} {et~al.}(2023){Somigliana}, {Testi}, {Rosotti}, {Toci}, {Lodato}, {Tabone}, {Manara}, \& {Tazzari}}]{somigliana2023}
{Somigliana}, A., {Testi}, L., {Rosotti}, G., {et~al.} 2023, \apjl, 954, L13

\bibitem[{{Somigliana} {et~al.}(2022){Somigliana}, {Toci}, {Rosotti}, {Lodato}, {Tazzari}, {Manara}, {Testi}, \& {Lepri}}]{somigliana2022}
{Somigliana}, A., {Toci}, C., {Rosotti}, G., {et~al.} 2022, \mnras, 514, 5927

\bibitem[{{Stadler} {et~al.}(2022){Stadler}, {G{\'a}rate}, {Pinilla}, {Lenz}, {Dullemond}, {Birnstiel}, \& {Stammler}}]{Stadler2022}
{Stadler}, J., {G{\'a}rate}, M., {Pinilla}, P., {et~al.} 2022, \aap, 668, A104

\bibitem[{{Stammler} \& {Birnstiel}(2022)}]{Stammler2022_Dustpy}
{Stammler}, S.~M., \& {Birnstiel}, T. 2022, \apj, 935, 35

\bibitem[{{Steinpilz} {et~al.}(2019){Steinpilz}, {Teiser}, \& {Wurm}}]{Steinpilz2019}
{Steinpilz}, T., {Teiser}, J., \& {Wurm}, G. 2019, \apj, 874, 60

\bibitem[{{Tabone} {et~al.}(2022){Tabone}, {Rosotti}, {Lodato}, {Armitage}, {Cridland}, \& {van Dishoeck}}]{Tabone2022}
{Tabone}, B., {Rosotti}, G.~P., {Lodato}, G., {et~al.} 2022, \mnras, 512, L74

\bibitem[{{Tabone} {et~al.}(2025){Tabone}, {Rosotti}, {Trapman}, {Pinilla}, {Pascucci}, {Somigliana}, {Alexander}, {Vioque}, {Anania}, {Kuznetsova}, {Zhang}, {P{\'e}rez}, {Cieza}, {Carpenter}, {Deng}, {Agurto-Gangas}, {Ruiz-Rodriguez}, {Sierra}, {Kurtovic}, {Miley}, {Gonz{\'a}lez-Ruilova}, {TorresVillanueva}, {Hogerheijde}, {Schwarz}, {Toci}, {Testi}, \& {Lodato}}]{AGEPRO_VII_population}
{Tabone}, B., {Rosotti}, G.~P., {Trapman}, L., {et~al.} 2025, \apj, 989, 7

\bibitem[{{Takeuchi} \& {Lin}(2002)}]{Takeuchi2002}
{Takeuchi}, T., \& {Lin}, D.~N.~C. 2002, \apj, 581, 1344

\bibitem[{{Tazzari} {et~al.}(2021){Tazzari}, {Testi}, {Natta}, {Williams}, {Ansdell}, {Carpenter}, {Facchini}, {Guidi}, {Hogherheijde}, {Manara}, {Miotello}, \& {van der Marel}}]{tazzari2021}
{Tazzari}, M., {Testi}, L., {Natta}, A., {et~al.} 2021, \mnras, 506, 5117

\bibitem[{{Testi} {et~al.}(2022){Testi}, {Natta}, {Manara}, {de Gregorio Monsalvo}, {Lodato}, {Lopez}, {Muzic}, {Pascucci}, {Sanchis}, {Miranda}, {Scholz}, {De Simone}, \& {Williams}}]{testi2022}
{Testi}, L., {Natta}, A., {Manara}, C.~F., {et~al.} 2022, \aap, 663, A98

\bibitem[{{Toci} {et~al.}(2023){Toci}, {Lodato}, {Livio}, {Rosotti}, \& {Trapman}}]{toci2023}
{Toci}, C., {Lodato}, G., {Livio}, F.~G., {Rosotti}, G., \& {Trapman}, L. 2023, \mnras, 518, L69

\bibitem[{{Trapman} {et~al.}(2025{\natexlab{a}}){Trapman}, {Zhang}, {Rosotti}, {Pinilla}, {Tabone}, {Pascucci}, {Agurto-Gangas}, {Anania}, {Carpenter}, {Cieza}, {Deng}, {Gonz{\'a}lez-Ruilova}, {Hogerheijde}, {Kurtovic}, {Kuznetsova}, {Miley}, {P{\'e}rez}, {Ruiz-Rodriguez}, {Schwarz}, {Sierra}, {TorresVillanueva}, \& {Vioque}}]{AGEPRO_V_gasmasses}
{Trapman}, L., {Zhang}, K., {Rosotti}, G.~P., {et~al.} 2025{\natexlab{a}}, \apj, 989, 5

\bibitem[{{Trapman} {et~al.}(2025{\natexlab{b}}){Trapman}, {Vioque}, {Kurtovic}, {Zhang}, {Rosotti}, {Pinilla}, {Carpenter}, {Cieza}, {Pascucci}, {Anania}, {Agurto-Gangas}, {Deng}, {Miley}, {P{\'e}rez}, {Sierra}, {Tabone}, {Ruiz-Rodriguez}, {Gonz{\'a}lez-Ruilova}, \& {TorresVillanueva}}]{AGEPRO_XI_gas_disk_sizes}
{Trapman}, L., {Vioque}, M., {Kurtovic}, N.~T., {et~al.} 2025{\natexlab{b}}, \apj, 989, 10

\bibitem[{{Tripathi} {et~al.}(2017){Tripathi}, {Andrews}, {Birnstiel}, \& {Wilner}}]{Tripathi2017}
{Tripathi}, A., {Andrews}, S.~M., {Birnstiel}, T., \& {Wilner}, D.~J. 2017, \apj, 845, 44

\bibitem[{{van der Marel} \& {Pinilla}(2023)}]{marelpinilla2023}
{van der Marel}, N., \& {Pinilla}, P. 2023, arXiv e-prints, arXiv:2310.09077

\bibitem[{{Villenave} {et~al.}(2020){Villenave}, {M{\'e}nard}, {Dent}, {Duch{\^e}ne}, {Stapelfeldt}, {Benisty}, {Boehler}, {van der Plas}, {Pinte}, {Telkamp}, {Wolff}, {Flores}, {Lesur}, {Louvet}, {Riols}, {Dougados}, {Williams}, \& {Padgett}}]{villenave2020}
{Villenave}, M., {M{\'e}nard}, F., {Dent}, W.~R.~F., {et~al.} 2020, \aap, 642, A164

\bibitem[{{Villenave} {et~al.}(2022){Villenave}, {Stapelfeldt}, {Duch{\^e}ne}, {M{\'e}nard}, {Lambrechts}, {Sierra}, {Flores}, {Dent}, {Wolff}, {Ribas}, {Benisty}, {Cuello}, \& {Pinte}}]{villenave2022}
{Villenave}, M., {Stapelfeldt}, K.~R., {Duch{\^e}ne}, G., {et~al.} 2022, \apj, 930, 11

\bibitem[{{Vioque} {et~al.}(2025){Vioque}, {Kurtovic}, {Trapman}, {Sierra}, {P{\'e}rez}, {Zhang}, {Curone}, {Rosotti}, {Carpenter}, {Tabone}, {Pinilla}, {Deng}, {Pascucci}, {Miley}, {Agurto-Gangas}, {Cieza}, {Anania}, {Ruiz-Rodriguez}, {Gonz{\'a}lez-Ruilova}, {TorresVillanueva}, \& {Kuznetsova}}]{AGEPRO_X_dust_disks}
{Vioque}, M., {Kurtovic}, N.~T., {Trapman}, L., {et~al.} 2025, \apj, 989, 9

\bibitem[{{Wada} {et~al.}(2011){Wada}, {Tanaka}, {Suyama}, {Kimura}, \& {Yamamoto}}]{Wada2011}
{Wada}, K., {Tanaka}, H., {Suyama}, T., {Kimura}, H., \& {Yamamoto}, T. 2011, \apj, 737, 36

\bibitem[{Warren \& Brandt(2008)}]{warren&brandt08}
Warren, S.~G., \& Brandt, R.~E. 2008, Journal of Geophysical Research: Atmospheres, 113, doi:https://doi.org/10.1029/2007JD009744.
\newblock \url{https://agupubs.onlinelibrary.wiley.com/doi/abs/10.1029/2007JD009744}

\bibitem[{{Weber} {et~al.}(2018){Weber}, {Ben{\'{\i}}tez-Llambay}, {Gressel}, {Krapp}, \& {Pessah}}]{Weber2018}
{Weber}, P., {Ben{\'{\i}}tez-Llambay}, P., {Gressel}, O., {Krapp}, L., \& {Pessah}, M.~E. 2018, \apj, 854, 153

\bibitem[{{Weidenschilling}(1977)}]{Weidenschilling1977}
{Weidenschilling}, S.~J. 1977, \mnras, 180, 57

\bibitem[{{Winter} {et~al.}(2018){Winter}, {Booth}, \& {Clarke}}]{winter2018}
{Winter}, A.~J., {Booth}, R.~A., \& {Clarke}, C.~J. 2018, \mnras, 479, 5522

\bibitem[{{Yen} {et~al.}(2017){Yen}, {Takakuwa}, {Chu}, {Hirano}, {Ho}, {Kanagawa}, {Lee}, {Liu}, {Liu}, {Matsumoto}, {Matsushita}, {Muto}, {Saigo}, {Tang}, {Trejo}, \& {Wu}}]{yen2017}
{Yen}, H.-W., {Takakuwa}, S., {Chu}, Y.-H., {et~al.} 2017, \aap, 608, A134

\bibitem[{{Zapata} {et~al.}(2020){Zapata}, {Rodr{\'\i}guez}, {Fern{\'a}ndez-L{\'o}pez}, {Palau}, {Estalella}, {Osorio}, {Anglada}, \& {Huelamo}}]{zapata2020}
{Zapata}, L.~A., {Rodr{\'\i}guez}, L.~F., {Fern{\'a}ndez-L{\'o}pez}, M., {et~al.} 2020, \apj, 896, 132

\bibitem[{{Zhang} {et~al.}(2025){Zhang}, {P{\'e}rez}, {Pascucci}, {Pinilla}, {Cieza}, {Carpenter}, {Trapman}, {Deng}, {Agurto-Gangas}, {Sierra}, {Kurtovic}, {Ruiz-Rodriguez}, {Vioque}, {Miley}, {Tabone}, {Gonz{\'a}lez-Ruilova}, {Anania}, {Rosotti}, {TorresVillanueva}, {Hogerheijde}, {Schwarz}, \& {Kuznetsova}}]{AGEPRO_I_overview}
{Zhang}, K., {P{\'e}rez}, L.~M., {Pascucci}, I., {et~al.} 2025, \apj, 989, 1

\bibitem[{{Zormpas} {et~al.}(2022){Zormpas}, {Birnstiel}, {Rosotti}, \& {Andrews}}]{zormpas2022}
{Zormpas}, A., {Birnstiel}, T., {Rosotti}, G.~P., \& {Andrews}, S.~M. 2022, \aap, 661, A66

\bibitem[{{Zubko} {et~al.}(1996){Zubko}, {Mennella}, {Colangeli}, \& {Bussoletti}}]{Zubko1996}
{Zubko}, V.~G., {Mennella}, V., {Colangeli}, L., \& {Bussoletti}, E. 1996, \mnras, 282, 1321

\bibitem[{{Zurlo} {et~al.}(2023){Zurlo}, {Gratton}, {P{\'e}rez}, \& {Cieza}}]{zurlo2023}
{Zurlo}, A., {Gratton}, R., {P{\'e}rez}, S., \& {Cieza}, L. 2023, European Physical Journal Plus, 138, 411

\end{thebibliography}

\appendix

\section{Synthetic observations: Comparison with CLEAN images} \label{app:comp_clean}

There are three main products to consider when generating an image from interferometric data with the CLEAN algorithm: a) the model image, which is generated from the visibilities; b) the residual image, which represents the Fourier Transform of the residual visibilities, whose amplitude should be comparable with the thermal noise background; and c) the point-spread-function (PSF) image, which shows the response of the visibility sampling to a point source emission. In order to estimate the angular resolution of an observation, a Gaussian function is fitted to the PSF image, thus quantifying the smallest possible angular scale that can be recovered in the image. Spatial variations in the model image that are smaller than the angular resolution could be the result of the reconstruction process, and do not necessarily represent an existing structure detected by the observation. Thus, the model image is convolved with the Gaussian representative of the PSF shape, erasing brightness structures of higher spatial frequency than those recoverable with the visibility coverage. As a final step, the convolved model is added with the residual image, and the result is the reconstructed image of CLEAN. 

In Sect.~\ref{section_Results}, properties such as total flux and disk size from radiative transfer models of the simulations have been compared to the AGE-PRO sources. In the case of radiative transfer models, those properties are easily obtained due to the absence of beam convolution or noise. However, for interferometric observations, they are more challenging to recover due to the limited angular resolution and sensitivity. Even though measurements with visibility modeling of the dust continuum emission contributes to increase the accuracy of results such as total flux and disk size, and partially overcoming the issues related to image reconstruction \citep[see][]{AGEPRO_X_dust_disks}, they are still limited to the physical constraints of the dataset, such as the signal-to-noise ratio at the longest baselines. 

For additional testing of the influence of the beam convolution and the thermal noise in the measured radii of the disk, we generated synthetic observations directly in the image plane, using the radiative transfer image as the CLEAN model \citep{hogbom1974, clark1980}, and convolving it with a Gaussian kernel representative of the angular resolution. We added a thermal noise image with a comparable standard deviation to that of the AGE-PRO observations. In comparison with the CLEAN algorithm, the main difference is in the method used to generate the underlying model. With CLEAN, the model is constructed as a collection of point sources and Gaussians, while in our simulated observations the model comes directly as an output of the radiative transfer. 

\begin{figure*}
\centering
\includegraphics[width=1.0\textwidth]{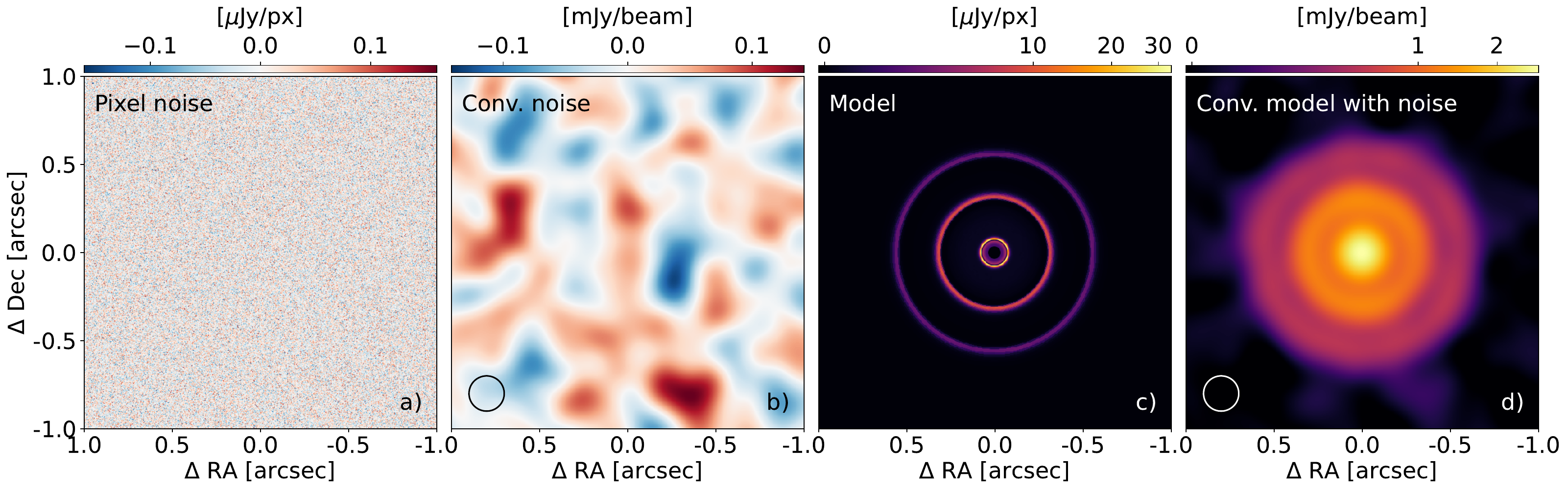}
\caption{Example of the different components of a synthetic observation. Panel a) shows an image of the pixel noise following a Gaussian distribution. Panel b) shows the same noise image, but convolved with a Gaussian of $0.25''$ of FWHM, and normalized to a standard deviation of $40\mu$Jy/beam. Panel c) shows the radiative transfer model at 1.3\,mm of one of the simulations. Panel d) shows the model convolved with the $0.25''$ Gaussian, after adding the convolved noise, mimicking the output image of the CLEAN algorithm. }
\label{fig:ex_thermal_noise}
\end{figure*}

The AGE-PRO observations have an average angular resolution of $0.25''$ and $0.8''$ in the 1.3\,mm and 1.05\,mm setups, and thus we use those values as the full width at half maximum of the Gaussian kernel to convolve the radiative transfer images. 
In the CLEAN algorithm, the sensitivity is set by the standard deviation of the intensity distribution in the residual map. In the idealized scenario where all the real emission is extracted in the CLEAN model, then the intensity of the CLEAN residuals should distribute as thermal noise. Thus, we represent the limited sensitivity in our synthetic observations by adding a thermal noise image, generated by assigning fluxes to each pixel following a Gaussian distribution centered at 0 and a standard deviation of $\mu=1$.  
An example of a thermal noise image is shown in the panel a) of Figure \ref{fig:ex_thermal_noise}. This thermal noise image is then convolved with the same beam as the radiative transfer image, and normalized to match a sensitivity of $25\,\mu\text{Jy}/\text{beam}$, comparable to AGE-PRO (see panel b) of Figure \ref{fig:ex_thermal_noise}). The convolved thermal noise image is then added to the convolved model image, thus obtaining a synthetic observation, as exampled in panel d) of Figure \ref{fig:ex_thermal_noise}. From these final images, we obtain the synthetic disk fluxes, sizes, and infer spectral indices. 

We calculated the $R_{90\%,cont}$ from the synthetic observations similarly to the AGE-PRO observations \citep{AGEPRO_XI_gas_disk_sizes}. The disk size is measured by calculating an azimuthally symmetric radial profile ($I_\nu:=I_\nu(r)$), and finding the radii that includes $90\%$ of the total flux. However, the measurement of the total flux is dependent on the size of the elliptical aperture where flux is integrated. The emission from the disks would only contribute positively to an intensity profile, but the contribution from thermal noise will oscillate between positive and negative values. Thus, the radius containing 100\% of a disk flux ($R_{100\%}$) is estimated from the radius at which the azimuthally averaged radial profile transitions from a positive value of $I_\nu$ to a negative value. The $R_{100\%}$ can also be interpreted as the radius at which the radial profile goes from being dominated by disk emission to a noise dominated regime. A equivalent method is to find the $R_{100\%}$ by looking at the cumulative flux radial profile of a disk, and finding the radius at which the first derivative of $I_\nu$ becomes 0. 

When measuring $R_{100\%}$ from the brightness profiles, an important caveat should be considered for disks with large depleted gaps or cavities, where the emission could become noise dominated before reaching the next emission substructure. In the synthetic images of our simulated disks, however, this is never the case, as there is no cavity or depleted gap larger than the beam size. Due to the relatively large beam compared to the disk sizes in the AGE-PRO observations, none of the cavities are resolved as empty either. 

\begin{figure}
\centering
\includegraphics[width=0.5\columnwidth]{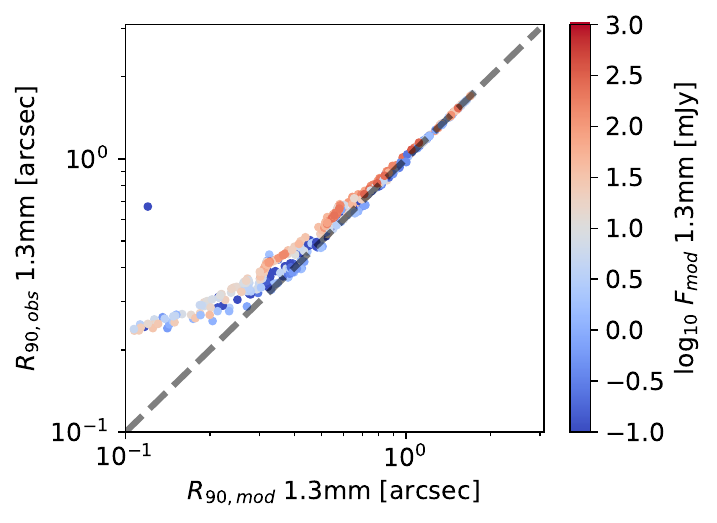}
 \caption{ Comparison between the $R_{90}$ in the radiative transfer images ($R_{90, mod}$)and in the synthetic observations ($R_{90, obs}$). Each dot represents a different disk, and images from all ages are included. The dots are colorcoded by the total flux of the model. }
 \label{fig:r90_r90}
\end{figure}

The synthetic images generated from the radiative transfer simulations do not have an inclination or rotation applied, and thus they simulate disks that have been observed in a face-on configuration. Since the beam convolution is done with a circular Gaussian kernel, there is no distortion of the flux in a preferred axis, as it is the case with interferometric observations, where there is typically one beam axis larger than the other. Thus, our measurements are not affected by either disk geometry or beam asymmetries. 
The measurement of disk sizes using the above described method proves robust for bright disks which are larger than approximately three times the angular resolution of the image (the FWHM of the beam in our synthetic observations is $0.250''$), as shown in Fig.~\ref{fig:r90_r90}.

\end{document}